\documentclass[10pt]{article}

\usepackage[a4paper,margin=21mm]{geometry}
\usepackage{amsmath,amssymb,amsthm,bm,mathtools}
\usepackage{graphicx,booktabs,multirow,array,makecell}
\usepackage{algorithm}
\usepackage{algpseudocode}
\usepackage[numbers,sort&compress]{natbib}
\usepackage{xcolor}
\usepackage{hyperref}
\usepackage{microtype}
\usepackage{enumitem}
\usepackage{caption}
\usepackage{placeins}
\usepackage{siunitx}
\usepackage{url}
\usepackage{hyperref}
\usepackage{hyphsubst}
\hypersetup{colorlinks=true,citecolor=blue,linkcolor=blue,urlcolor=blue}
\graphicspath{{Figures/}}
\newcommand{\email}[1]{\protect\href{mailto:#1}{#1}}
\newcommand{\R}{\mathbb{R}}
\newcommand{\C}{\mathbb{C}}
\newcommand{\J}{\mathbf J}

\newcommand{\Hankel}{\mathcal H}
\newcommand{\Proj}{\mathcal P}
\newcommand{\Model}{\mathbb{M}_{\hat\Bp}}
\newcommand{\norm}[1]{\left\lVert #1\right\rVert}

\newtheorem{problem}{Problem}

\def\Bk{{\bf k}}
\def\Bb{{\bf b}}
\def\Bc{{\bf c}}

\def\Bp{{\bf p}}

\newcommand{\RR}{\mathbb{R}}
\newcommand{\ds}{\displaystyle}
\def\bx{{{\bf x}}}

\def\hbx{{\hat{\bx}}}
\def\bz{{{\bf z}}}
\def\by{{{\bf y}}}
\def\bel{{\boldsymbol{\ell}}}
\def\Eta{{\boldsymbol{\eta}}}

\def\BJ{{\bf J}}
\def\tJ{\widetilde{\J}}
\def\BE{{\bf E}}
\def\BH{{\bf H}}

\def\BA{{\bf A}}
\def\BY{{\bf Y}}

\begin{document}
    \title{Model-Consistent Structured-Hankel Matrix Completion for 3D Sparse Multi-frequency Electromagnetic Source Reconstruction}
    \author{
        Shujaat Khan\footnotemark[2] \footnotemark[3]
		\and
		Abdul Wahab\footnotemark[1] \footnotemark[4] 
	}
	\maketitle
	
	\renewcommand{\thefootnote}{\fnsymbol{footnote}}
    \footnotetext[1]{Corresponding author.}
    \footnotetext[2]{Department of Computer Engineering, College of Computing and Mathematics, King Fahd University of Petroleum \& Minerals, Dhahran 31261, Saudi Arabia (\email{shujaat.khan@kfupm.edu.sa}).}
    \footnotetext[3]{ SDAIA-KFUPM Joint Research Center for Artificial Intelligence, King Fahd University of Petroleum and Minerals, Dhahran 31261, Saudi Arabia.}
    \footnotetext[4]{Department of Mathematics, College of Science, Sultan Qaboos University, Muscat 123, Oman (\email{a.wahab@squ.edu.om}).}
	\renewcommand{\thefootnote}{\arabic{footnote}}
    
\begin{abstract}
Reconstruction of three-dimensional (3D) electromagnetic currents from sparse, multi-frequency, far-field radiation data is severely ill-posed because sparse sampling masks specific current Fourier modes and the missing spectrum enlarges the effective null space of the forward operator. Although structured-Hankel matrix completion can recover missing spectrum by exploiting the finite rate of innovations structure of a compactly supported geometrically sparse source, it treats current components independently without guaranteeing compatibility with the model. This often renders non-physical solutions that violate model consistency. To resolve this, we propose a training-free Maxwell-model-consistent 3D structured-Hankel framework (MC-Hankel) that alternates joint low-rank Hankel completion with an exact, closed-form projection onto the true Maxwell synthesis subspace while enforcing conjugate symmetry and noise-aware data fidelity. Across tests (with five paired randomized trials) on two source models (curl and mixed-type currents), three sparse sampling rates (30\%, 40\%, and 50\% of a Nyquist-sampled grid), and clean versus noisy (10 dB signal-to-noise ratio (SNR) additive white Gaussian noise) conditions, MC-Hankel consistently outperforms sub-sampled Fourier inversion, scale-normalized $\ell_1$-compressed sensing baseline, and the standard joint 3D annihilating filter-based low-rank Hankel matrix completion approach (ALOHA). Across all considered source-noise-sampling configurations, MC-Hankel improves mean peak SNR (PSNR) by $0.65$--$4.22$ dB and tri-planer structural similarity (3D SSIM) by $0.0266$--$0.0703$, reduces relative full-volume $\ell_2$ reconstruction error by $7.4\%$--$38.9\%$, and brings non-physical model residuals down to machine precision over standard joint 3D ALOHA with merely $17\%$ computational overhead. Exact signed-rank tests are underpowered for five paired trials, so statistical conclusions are reported together with paired bootstrap intervals, effect sizes, and small-sample limitations.

\end{abstract}


\section{Introduction}

Electromagnetic inverse source problems (ISPs) seek spatial current distributions from the radiation pattern of the electric or magnetic fields outside the current support. They arise in antenna synthesis \cite{LeoneMaistoPierri2018}, biomedical imaging and diagnostics \cite{Thio23, Beltrachini21, Jerbi04}, nondestructive evaluation of structural health \cite{Takahashi22}, security robotics \cite{AmmariN}, and electromagnetic imaging \cite{AmmariBaoFleming2002, Albanese06}. The ISPs are severely ill-posed due to the presence of non-radiating sources, especially when the apparatus does not permit full aperture acquisition \cite{ISPs, Valdivia12, WAHAB}. Limited aperture and sparse acquisition mask specific spatial frequency information, causing components of the unknown source to behave like non-radiating sources. This further enlarges the effective null space of the forward (i.e., source-to-data) operator, as the masked components reside there, making the inverse operator severely ill-posed  \cite{Bleistein}. Nyquist-sampled grid multi-frequency acquisition alleviates this difficulty by sampling different spatial Fourier modes of the source, but such dense coverage of the temporal frequency and observation direction is often impractical.

Fourier inversion provides a direct route from admissible multi-frequency far-field samples to the coefficients of a polarization-based transverse electric (TE)-transverse magnetic (TM) source decomposition \cite{WangEM2019}. When the current density is spatially compact and geometrically sparse or piecewise regular, its spectrum has a finite-rate-of-innovations (FRI) structure \cite{vetterli2002sampling}. Under FRI signal theory, the geometric properties of the source are encapsulated within a sparse set of low-frequency spatial modes. Consequently, the associated multichannel Hankel-lifting of a concatenated spectral vector is low-rank under suitable parametric choices. It provides an impetus behind framing a matrix completion problem under a low-rank constraint for data enrichment seeking spectral completion for unregistered Fourier modes. Such matrix completion can be effectively done, for instance, by using the annihilating filter-based low-rank Hankel matrix completion approach (ALOHA) and related structured spectral recovery methods \cite{ChenChi2014, JinYe2015, JinLeeYe2016, JacobManiYe2020}. The spectral completion strategy for acoustic and electromagnetic source reconstructions using sparse multi-frequency data was applied in \cite{GuoWahabWang2023, arxiv}. Recent works have improved the computational and numerical behavior of structured low-rank recovery through robust structured gradient descent \cite{CaiCaiYou2023}, convex multichannel formulations \cite{WuYangXu2024}, Newton-like updates \cite{CaiHuangLuYou2025}, and multi-measurement Hankel tensor completion \cite{LiZhangWuCui2025}. In parallel, learned inverse source and operator-learning methods have emerged for source localization \cite{WiLeeOllerFazeli2025, ChenChangGuoWang2026} and sparse-data inversion \cite{DongSuLiuChenChen2026}. Sparse phased and phaseless formulations \cite{JiLiu2020}, factorization methods \cite{GriesmaierSchmiedecke2017}, and multiscale recovery \cite{LiLiu2023} have subsequently expanded the available theory and reconstruction tools. 

In the multichannel Hankel lifting for vector fields like current source density, a joint 3D Hankel matrix is formed using a spectral vector concatenating those of the constituent TE/TM source components obtained through polarization-based Helmholtz-Hodge decomposition of the current density \cite{arxiv}. One limitation of the standard joint 3D ALOHA or related structured spectral recovery methods is that they treat constituent source components independently without guaranteeing compatibility with the underlying model. A completion recovers coefficients for each Fourier mode that are generated by paired scalar fields through their shared low-rank Hankel representation without enforcing model-specific physical constraints. This often produces non-physical and inconsistent sources. Specifically, a spectral low-rank solution can agree with the observed data while rendering nonphysical components orthogonal to the synthesis subspace corresponding to the Maxwell equations and the source model. 

In this work, we propose a training-free Maxwell-model-consistent 3D structured-Hankel framework (MC-Hankel) that alternates joint low-rank Hankel completion with an exact, closed-form projection onto the true Maxwell synthesis subspace while enforcing conjugate symmetry and noise-aware data consistency. In particular, the present work has four main contributions.

\begin{enumerate}[leftmargin=6mm]
\item A joint 3D structured-Hankel formulation for all current components, augmented by an explicit per-frequency Maxwell source-subspace constraint.

\item A closed-form orthogonal projection that estimates the Fourier coefficients of the constituent TE/TM scalar densities consistent with the polarization-based Helmholtz-Hodge decomposition of the unknown current density.

\item An alternating algorithm that combines low-rank completion, Maxwell-model projection, conjugate symmetry, and noise-aware data consistency, while keeping the Hankel rank, patch, penalty, and iteration budget matched to standard joint 3D ALOHA.

\item A full-volume paired evaluation with 2D slices, 3D isosurfaces, error projections, topology metrics, runtimes, bootstrap intervals, effect sizes, and an explicit small-sample significance analysis.
\end{enumerate}

The novelty of our work lies in its model-specific approach, rather than in proposing a new generic theory of low-rank completion; specifically, we derive an exact Maxwell model current projector and embed it into each iteration of joint 3D Hankel completion for sparse multi-frequency inverse source reconstruction. 

The structure of the paper is as follows. In section \ref{s:Pre}, we provide a mathematical formulation of the electromagnetic ISP, Fourier inversion algorithm, and structured-Hankel matrix completion framework. In section \ref{s:Num}, we provide details of the experimental setup, comparison methods, evaluation metrics, and statistical analysis. In section \ref{s:results}, we present substantive numerical simulations with quantitative and qualitative comparisons, statistical analysis, and limitations of MC-Hankel. The research is summarized in section \ref{s:Conc}.

\section{Electromagnetic ISP and Hankel matrix completion}\label{s:Pre}
In this section, we mathematically formulate the electromagnetic ISP with sparse multi-frequency far-field data and frame a model-consistent low-rank structured-Hankel matrix completion problem for sparse data enrichment and reconstructing electric current source density.

\subsection{Electromagnetic inverse source formulation}
Let $\R^3$ be a homogeneous isotropic electromagnetic medium characterized by an electric permittivity $\varepsilon_0\in\R_+$ and magnetic permeability $\mu_0\in\R_+$. Let $\J:\R^3\to\R^3$ be a smooth electric current source density, compactly supported in \(D=(-a/2,a/2)^3\subset\R^3\), for $a\in\R_+$. Let $\BE:\RR^3\times\R_+\to\RR^3$ and $\BH:\RR^3\times\R_+\to\RR^3$ be the time-harmonic electric and magnetic fields radiated by the current source density $\BJ$, satisfying the time-harmonic Maxwell equations, 
\begin{align}
 i\omega\varepsilon_0\BE(\bx,\omega)+\nabla\times\BH(\bx,\omega)=\BJ(\bx) 
 \text{ and } 
 -i\omega\mu_0\BH(\bx,\omega) +\nabla\times\BE(\bx,\omega)=\mathbf{0}, \quad\text{for }\bx\in\R^3\text{ and }i=\sqrt{-1},\label{ME}
\end{align}
subject to the Silver-M\"uller radiation conditions,
\begin{align}
    \lim_{|\bx|\to\infty} \left(
    \sqrt{\mu_0} \BH(\hbx,\omega) \times \hbx -\sqrt{\varepsilon_0} \BE(\hbx,\omega)\right)=\mathbf{0}=\lim_{|\bx|\to\infty}
    \left(\sqrt{\varepsilon_0} \BE(\hbx,\omega) \times \hbx +\sqrt{\mu_0} \BH(\hbx,\omega)\right),\label{RC}
\end{align}
where $\bx\neq \mathbf{0}$,  $\hat{\bx}:={\bx}/{|\bx|}\in\left\{\bx\in\RR^3:\,|\bx|=1\right\}=:\mathbb{S}^2$, and $\omega\in\R_+$ is the angular frequency. Since the electric and magnetic formulations are mathematically similar, we proceed with the magnetic formulation of the Maxwell equations \eqref{ME} for the remainder of this work and assume the aperture acquisition of the magnetic far-field data. To that end, we eliminate $\BE$ from \eqref{ME} to get the uncoupled vector Helmholtz equation in $\BH$, 
\begin{equation}
\nabla\times\nabla\times\BH(\bx,\omega)-\kappa^2\BH(\bx,\omega)
=\nabla\times \J(\mathbf x),
\qquad \kappa:=\omega\sqrt{\mu_0\varepsilon_0}, \quad \bx\in\RR^3.
\label{eq:wave}
\end{equation}
Thanks to Silver-M\"{u}ller radiation conditions \eqref{RC}, $\BH$ behaves smoothly away from the source $\J$. 
 Specifically, there exists an analytic function $\BH_\infty:\mathbb{S}^2\to \RR^3$, called the magnetic far-field \cite{Colton-Kress}, such that 
$$
\BH(\bx,\omega)=\dfrac{e^{i\kappa|\bx|}}{|\bx|}\left\{\BH_\infty(\hat{\bx},\omega)+O\left(\dfrac{1}{|\bx|}\right)\right\}, \qquad\text{as }|\bx|\to +\infty.
$$ 

Let $\bel:=(\ell_1,\ell_2,\ell_3)\in\left\{-N,\cdots,N\right\}^3=:\mathbb{S}_N$ be a multi-index, for $N\in\mathbb{N}$, and  $\epsilon_0\in\R_+$ be a small constant such that $\epsilon_0\to 0^+$. We define admissible sampling points $\hbx_\bel:=\bel/|\bel|$ and wavenumbers $\kappa_\bel:=2\pi|\bel|/a$ for $\bel\neq \mathbf{0}$. We set $\hbx_{\mathbf{0}}:=(1,0,0)$ and $\kappa_\mathbf{0}:=2\pi\epsilon_0/a$. We call a polarization direction $\hat\Bp$ admissible if $\hat\Bp\times\bel\neq \mathbf{0}$, for all $\bel\in\mathbb{Z}^3$. For such polarization directions, $\BJ$ admits a \text{Helmholtz-Hodge decomposition} \cite{Lindell}, 
\begin{equation}
    \BJ(\bx) = f(\bx)\hat{\Bp} +\hat{\Bp} \times \nabla g(\bx), \label{Helm-Decomp}
\end{equation}
where  $f:\RR^3 \to \RR$ and $g:\RR^3\to \RR$ are the TE and TM scalar components, respectively. 

In this work, we address the following ISP. 

\begin{problem}[Sparse-Data Electromagnetic ISP]\label{prob:ISP}
Let $\Omega\subset \mathbb{S}_N$ be a sparse multi-index set (in the sense that $|\Omega|\ll|\mathbb{S}_N|)$ containing $\bel=\mathbf{0}$ such that if $\bel\in\Omega$ then $-\bel\in\Omega$ and $\hat{\Bp}$ be an admissible polarization direction. Let $\BH$ be the magnetic field satisfying \eqref{eq:wave}. Find the radiating current $\BJ$ using the sparse multi-frequency far-field dataset
$$
\left\{\BH_\infty(\hat{\bx}_\bel, \kappa_{\bel})\,\left|\right.\,\,\bel\in\Omega\subset\mathbb{S}_N\right\}.
$$
\end{problem}

\subsection{Fourier inversion algorithm}
The magnetic far-field patterns are proportional to directional samples of the Fourier series of \(\BJ\); hence, admissible observation direction-wavenumber pairs $(\hat{\bx}_\bel, \kappa_{\bel})$ provide source Fourier coefficients \cite{WangEM2019}. Indeed, if $\varphi_{\bel}$ and  $(f_\bel,g_\bel)$ are the Fourier basis functions and corresponding Fourier coefficients of $(f,g)$, respectively, i.e., 
$$
\varphi_{\bel}(\bx):=\ds\exp\left(\dfrac{2\pi i}{a}\hat{\bx}_{\bel}^T\bx\right), 
\qquad
{f}_\bel:=\dfrac{1}{a^3}\int_{D}f(\bx)\overline{\varphi_\bel(\bx)}d\bx,
\qquad
{g}_\bel:=\ds\dfrac{1}{a^3}\int_{D}g(\bx)\overline{\varphi_\bel(\bx)}d\bx, 
\qquad
\bel\in\mathbb{Z}^3,
$$
then 
\begin{equation}
    \BJ(\bx)\approx\hat{\Bp} f_\mathbf{0}+\sum_{1\leq \|\bel\|_\infty \leq N}\left(\hat{\Bp} f_{\bel}+\dfrac{2\pi i}{a}(\hat{\Bp}\times \bel)g_{\bel}\right)\varphi_{\bel}(\bx),\label{FM}
\end{equation}
where 
\begin{align}
    f_\bel&= \dfrac{4\pi\hat{\bx}_\bel\times\hat{\Bp}\cdot\BH_\infty(\hat{\bx}_\bel,\kappa_\bel)}{i\kappa_\bel a^3 |\hat{\bx}_\bel \times \hat{\Bp}|^2} \quad\text{and}\quad
    g_\bel=- \dfrac{2\hat{\bx}_\bel\times(\hat{\Bp}\times \bel)\cdot\BH_\infty(\hat{\bx}_\bel,\kappa_\bel)}{\kappa_\bel a^2 |\hat{\bx}_\bel \times (\hat{\Bp}\times \bel)|^2},\qquad \bel\in\mathbb{S}_N,\, \bel\neq\mathbf{0},\label{flh}
    \\
    f_{\mathbf{0}}&\approx \dfrac{\epsilon_0\pi}{a^3\sin (\epsilon_0\pi)}\left(\dfrac{4\pi\hbx_{\mathbf{0}}\times\hat{\Bp}\cdot\BH_\infty(\hbx_{\mathbf{0}},\kappa_{\mathbf{0}})}{i\kappa_{\mathbf{0}}|\hbx_{\mathbf{0}}\times\hat{\Bp}|^2}-\sum_{1\leq \|\bel\|_\infty \leq N} f_\bel \int_D\exp\left(i(\kappa_\bel\hat{\bx}_{\bel}^T-\kappa_{\mathbf{0}}\hbx_{\mathbf{0}}^T)\by\right)d\by\right).\label{f0lh}
\end{align}
Here, superposed bar represents complex conjugate and $\norm{\bel}_\infty:=\max_{1\leq k\leq 3}|\ell_k|$, is $\ell_\infty$-norm.
Note that the Fourier coefficient vector, $\tJ_\bel$, of the current $\BJ$ has the form 
\begin{equation}
\tJ_{\bel}=f_\bel\hat\Bp +g_\bel\Bb_\bel =\BA_\bel\Bc_\bel, \qquad\bel\in\mathbb{S}_N,
\label{eq:fourier_model}
\end{equation}
where
\begin{equation}
\Bb_\bel:=i(\hat\Bp\times\Bk_\bel), \qquad
\Bk_\bel:=\dfrac{2\pi}{a}\bel, \qquad
\Bc_\bel:=\begin{bmatrix} f_\bel & g_\bel\end{bmatrix}^{T}, \qquad
\BA_\bel:=\begin{bmatrix}\hat\Bp&\Bb_\bel\end{bmatrix}, \qquad
\bel\in\mathbb{S}_N,\, \bel\neq\mathbf{0}.
\label{eq:A}
\end{equation}
Let \(\Proj_{\Omega}\) be the coefficient selection (\textit{mask}) over the multi-index set $\Omega$. Then Problem \ref{prob:ISP} can be recast as a spectrum recovery problem below.

\begin{problem}[Spectrum Recovery Problem]
\label{prob:Spectrum}
Find $\tJ\in\C^{3n}$ satisfying \eqref{eq:fourier_model} from the sparse measurements $\BY\in\C^{3n}$ given by 
\begin{equation}
\BY:=\Proj_{\Omega}(\tJ)+\Eta,
\label{eq:data}
\end{equation}
where $n:=2N+1$, and 
$
\Eta\sim\mathcal{CN}\left(\mathbf{0},{P_{\BH}}/{10^{\mathrm{SNR}/10}}\textbf{I}\right) \in\C^{3n}$ is circular complex additive white Gaussian noise (AWGN). Here, $\Eta=\mathbf{0}$ for clean data, $P_{\BH}$ is the mean squared magnitude of the clean far-field data, and $\mathbf{I}$ is identity matrix.
\end{problem}

\subsection{Structured-Hankel reconstruction}\label{s:HankelRec}
A structured-Hankel matrix completion approach aims at solving Problem \ref{prob:Spectrum} through a Hankel-lifting of the spectral vector $\tJ$ exploiting its FRI property \cite{vetterli2002sampling}. As $\BJ$ is geometrically sparse and compactly supported, its geometric information is encapsulated within a sparse set of low-frequency spatial modes. Consequently, the associated Hankel-lifting of $\tJ$ is low-rank. Below, we provide details of the structured-Hankel completion approach and the proposed MC-Hankel. 

\subsubsection{Joint 3D Hankel-lifting}
For a generic vector $\mathbf{v}\in\RR^n$, a wrap-around structured-Hankel matrix with patch-size $p<n$ is defined as
$$
\mathcal{H}_p(\mathbf{v}) = \begin{bmatrix}  v_1 & v_2 & \dots & v_p \\ v_2 & v_3 & \dots & v_{p+1} \\ \vdots & \vdots & \ddots & \vdots \\ v_n & v_1 & \dots & v_{p-1} \end{bmatrix} \in \mathbb{R}^{n \times p}.
$$
For a scalar cube \(X\in\C^{n\times n\times n}\), let \(\Hankel_p(X)\) stack all overlapping \(p\times p\times p\) neighborhoods into a multichannel Hankel matrix. 
Let $(\tJ_{\bel})_k$ be the $k$-th component of $\tJ_\bel$ and $\tJ_k:=[(\tJ_{\bel})_k]\in\C^{n}$, for $k=1,2,3$. Then, the joint vector-current Hankel-lifting of $\tJ$ is defined as
\begin{equation}
\Hankel_{\mathrm{joint}}(\tJ)
=\begin{bmatrix}
\Hankel_p(\tJ_1) &
\Hankel_p(\tJ_2) &
\Hankel_p(\tJ_3)
\end{bmatrix}.
\label{eq:joint_hankel}
\end{equation}

The joint Hankel completion recasts Problem \ref{prob:Spectrum} and solves the rank-minimization problem subject to data fidelity constraint,
\begin{equation*}
\ds\arg\min _{\tJ} \left(\text{rank}(\Hankel_{\mathrm{joint}}(\tJ))\right)  
\quad\text{s.t.}\quad \Proj_{\Omega}(\tJ)=\BY,
\end{equation*}
through its standard convex relaxation, 
\begin{equation*}
\ds\arg\min_{\tJ} \norm{\Hankel_{\mathrm{joint}}(\tJ)}_*
\quad\text{s.t.}\quad \Proj_{\Omega}(\tJ)=\BY.
\end{equation*}
The standard joint 3D ALOHA further parametrizes the matrix nuclear norm $\norm{\cdot}_*$ through a fixed-rank factorized matrix form \cite{ye2016compressive, FazelPongSunTseng2013} and solves constrained optimization problem,
\begin{equation}
\arg\min_{\tJ,\mathbf{U},\mathbf{V}}
\frac{1}{2}\left(\norm{\mathbf{U}}_F^2+\norm{\mathbf{V}}_F^2\right)
\quad\text{s.t.}\quad
\Hankel_{\mathrm{joint}}(\tJ)=\mathbf{U}\mathbf{V}^H,
\quad
\Proj_{\Omega}(\tJ)=\BY,
\label{eq:aloha}
\end{equation}
where $\norm{\cdot}_F$ is the Frobenius norm and superposed $H$ is the Hermitian transpose. The constrained optimization problem \eqref{eq:aloha} is then solved using \textit{alternating-direction method of multipliers (ADMM) }\cite{ye2016compressive}.

\subsubsection{Model-consistent Hankel-lifting}

The minimization problem \eqref{eq:aloha} exploits common spatial support among the current components but does not enforce source model compatibility \eqref{eq:fourier_model}. Therefore, a \textit{minimum-rank} solution may have data fidelity but may not be compatible with the current source model. To resolve this, we seek a minimum-rank solution with data fidelity in the model synthesis subspace,
\begin{equation}
    \Model:=\left\{\tJ\in\C^{3n}\left|\right.\tJ_\bel=f_\bel\hat\Bp +g_\bel\Bb_\bel, \,\,\bel\in\mathbb{S}_N\right\},
\end{equation}
for a known admissible polarization direction $\hat\Bp$. To that end, we define a local orthogonal projection operator on the subspace $\Model$ as
\begin{equation}
\Proj_{\Model}(\bz_{\bel})
:= \BA_{\bel}
\left(\BA_{\bel}^{H}\BA_{\bel}\right)^{\dagger}
\BA_{\bel}^{H}\bz_{\bel},\quad \bz_{\bel}\in\C^3, \,\bel\in\mathbb{S}_N,
\label{eq:projector}
\end{equation}
where \(\dagger\) denotes the Moore-Penrose inverse and $\BA_\bel$ is defined in \eqref{eq:A}. 

By definition of $\Bb_\bel$ in \eqref{eq:A}, \(\hat\Bp^H\Bb_{\bel}=0\), for all $\bel\in\mathbb{S}_N$ and polarization directions $\hat\Bp\in\R^3$. Consequently, when \(\Bb_{\bel}\neq\mathbf{0}\), we have 
\begin{align*}
f_\bel=\frac{\hat\Bp^H\bz_\bel}{\norm{\hat\Bp}_2^2},
\qquad
g_\bel=\frac{\mathbf b_\bel^H\bz_\bel}{\norm{\Bb_\bel}_2^2},\qquad \bz_{\bel}\in\C^3, \,\bel\in\mathbb{S}_N,\,\bel\neq \mathbf{0}.
\end{align*}
Note that, only the component parallel to an admissible direction \(\hat\Bp\) is identifiable when \(\Bb_\mathbf{0}=\mathbf{0}\). Also, the projector $\Proj_{\Model}$ only retains $g_\bel$ component, i.e.,
\begin{equation}
\Proj_{\Model}(\bz_\bel)
=\left(\dfrac{\Bb_\bel^{H}\mathbf z_\bel}{\norm{\Bb_\bel}_2^2}\right)\Bb_\bel, \qquad\bz_\bel\in\C^3,\,\bel\in\mathbb{S}_N, \, \bel\neq\mathbf{0},
\end{equation}
for curl-type sources and sets the origin to zero.

The \textit{Maxwell-model projector}, \(\Proj_{\Model}\), applies the appropriate projector independently at every Fourier mode. Specifically, a relaxation parameter \(\gamma\in[0,1]\) permits an update
\begin{equation}
\tJ_\bel^{\mathrm{proj}}
:=(1-\gamma)\tJ_\bel+\gamma\Proj_{\Model}(\tJ_\bel), \quad\bel\in\mathbb{S}_N,
\label{eq:relaxation}
\end{equation}
so that \(\gamma=1\) for an exact model consistency. 

The proposed MC-Hankel method solves the constrained optimization problem
\begin{equation}
\arg\min_{\tJ\in\Model}\ \norm{\Hankel_{\mathrm{joint}}(\tJ)}_{*}
\quad\text{s.t.}\quad
\Proj_{\Omega}(\tJ)\approx\BY,
\quad
\tJ_{-\bel}=\overline{\tJ_\bel}.
\label{eq:proposed_problem}
\end{equation} 
The last conjugate-symmetry constraint follows from a real-valued spatial current. Similar to standard joint 3D ALOHA, the implementation of the MC-Hankel uses the parametrization of the nuclear norm through a fixed-rank factor matrix form and solves the optimization problem
\begin{equation}
\arg\min_{\tJ\in\Model,\mathbf{U},\mathbf{V}}
\frac{1}{2}\left(\norm{\mathbf{U}}_F^2+\norm{\mathbf{V}}_F^2\right)
\quad\text{s.t.}\quad
\Hankel_{\mathrm{joint}}(\tJ)=\mathbf{U}\mathbf{V}^H,
\quad
\Proj_{\Omega}(\tJ)=\BY, \quad \tJ_{-\bel}=\overline{\tJ_{\bel}},
\label{eq:MC-Hankel}
\end{equation}
using ADMM-style updates. After \textit{Hankel-unlifting}, each iterate is projected onto \(\Model\), symmetrized, and blended with model-projected observed coefficients,
\begin{equation}
\tJ^{k+1}
=\alpha\,\Proj_{\Model}\circ\Proj_\Omega(\tJ^k)
+(1-\alpha)\,\tJ^{k},
\label{eq:data_consistency}
\end{equation}
where the data-weight $\alpha\in[0,1]$, defined by
$$
\alpha:=1-\dfrac{1}{\sqrt{10^{\mathrm{SNR}/10}+1}},
$$
handles the awareness of the iterates to the measurement noise $\Eta$ with \(\alpha=1\) for clean experiments.

\begin{algorithm}[!htb]
\caption{Maxwell-model-consistent joint 3D Hankel completion (MC-Hankel)}
\label{alg:mc_hankel}
\begin{algorithmic}[1]
\Require Sparse vector coefficients \(\BY\), mask \(\Omega\), polarization $\hat\Bp$, patch size \(p\), target rank \(r\), penalty \(\mu\), iterations \(K\), data-weight \(\alpha\), relaxation \(\gamma\)  
\State Project and symmetrize the measured coefficients; normalize by their maximum observed magnitude
\State Form \(\Hankel_{\mathrm{joint}}(\BY)\) and initialize rank-\(r\) factors by truncated SVD
\For{$k=1,\ldots,K$}
    \State Hankel-unlift the current factorized estimate
    \State Apply the relaxed Maxwell projection in Eq.~\eqref{eq:relaxation}
    \State Enforce conjugate symmetry
    \State Enforce noise-aware data fidelity using Eq.~\eqref{eq:data_consistency}
    \State Hankel-lift and update the low-rank factors and dual variable
\EndFor
\State Apply a final model projection, conjugate-symmetry projection, and data-consistency update
\State Rescale and synthesize the 3D spatial current
\end{algorithmic}
\end{algorithm}

\section{Experiments setup and configuration}\label{s:Num}
Details of the experiential setup and configuration for comparison methods, evaluation metrics, and statistical analysis are provided below. A summary of the experimental configuration is provided in Table \ref{tab:configuration}.

\begin{table}[!htb]
\centering
\caption{Experimental configuration used for all reported results.}
\label{tab:configuration}
\small
\begin{tabular}{p{0.34\linewidth}p{0.56\linewidth}}
\toprule
Item & Setting \\
\midrule
Domain and Fourier grid & $D=(-1/2,1/2)^3$, $N=10$, $n=21^3=9261$ coefficient locations \\
Source models & J1: pure curl; J2: mixed longitudinal and curl source \\
Reference/source grids & $50^3$ stored source samples; $101^3$ Fourier-synthesis/evaluation grid \\
Sampling & Nested conjugate-paired masks at $30\%$, $40\%$, and $50\%$ \\
Noise & No noise and circular complex AWGN at $10$ dB \\
Trials & Five independent paired trials; common mask/noise within each trial \\
$\ell_1$-CS & Complex 3D DCT; $\rho=20$; 800 iterations; tolerance $10^{-6}$ \\
Hankel parameters & Patch $3\times\times3$; rank 22 (J1), rank 30 (J2); $\mu=10$; 30 iterations \\
Model projection & Exact, $\gamma=1$; pure-curl constraint $f=0$ \\
Data weight & $\alpha=1$ for clean data; $\alpha=0.6985$ for $10$-dB AWGN \\
Primary evaluation & Full-volume PSNR, Rel.$L_2$, tri-planar SSIM, Dice$_{25/55}$, model residual, runtime \\
Statistics & Exact paired Wilcoxon, Holm correction, 10,000 paired bootstrap samples, Cohen's $d_z$, win rate \\
\bottomrule
\end{tabular}
\end{table}

\subsection{Data, source models, and spatial sampling grids}

The numerical simulations in the next section use two precomputed 3D Fourier-Maxwell datasets generated on \(D=(-1/2,1/2)^3\), i.e., \(a=1\). The Fourier truncation order is set to \(N=10\), rendering \(n=(2N+1)^3=21^3=9261\) coefficient locations, including the origin. Far-field data are converted to scalar coefficients through the Fourier formulas \eqref{flh}-\eqref{f0lh} and then assembled into vector coefficients in Eq.~\eqref{eq:fourier_model}.

We consider two source configurations, referred to as J1 and J2. 
\begin{itemize}
    \item Source model J1: a smooth curl-type source of the form 
    \(\J=\hat\Bp\times\nabla g\) with
$$
\hat\Bp=\begin{bmatrix} 0.559017 &-0.500000 & 0.661438\end{bmatrix}^T.
$$
    \item Source model J2: a mixed source of the form \(\J=\hat\Bp f+\hat\Bp\times\nabla g\) with 
    $$
    \hat\Bp=\begin{bmatrix} 0.745356 &-0.333333 &0.577350\end{bmatrix}^T.
    $$
\end{itemize}
The reference source samples are stored on a \(50\times 50\times 50\) grid, while Fourier synthesis for all quantitative evaluations is performed on a common \(101\times101\times101\) spatial grid. The fully sampled band-limited reconstruction is used as the numerical reference, so the comparison measures degradation caused by sparse sampling, noise, and completion rather than Fourier truncation error.

\subsection{Sampling rates, noise, and paired trials}

Nested conjugate-paired masks retain nominally \(30\%\), \(40\%\), and \(50\%\) of the coefficient lattice. The origin is included, and if a mode is retained, then its conjugate counterpart is also retained. Within each trial, every method receives the same mask and the same noisy coefficients. Across trials, masks are independently regenerated; noisy experiments also use independent AWGN realizations. Five paired trials were completed for every source, noise condition, and sampling rate.

The noisy condition adds circular complex AWGN at \(10\) dB SNR relative to the mean complex far-field signal power.
For \(10\) dB AWGN, the code uses
$$\alpha=1-\frac{1}{\sqrt{10^{10/10}+1}}\approx 0.6985,$$ 
bounded below by \(0.65\). Clean and noisy datasets are otherwise processed identically. Deterministic seeds are recorded by the code to reproduce every realization. 

\subsection{Compared methods and their configurations}
The following methods are evaluated:
\begin{enumerate}[leftmargin=6mm]
\item \textbf{Sub-sampled Fourier inversion:} zero-filling of masked coefficients followed by Fourier synthesis.
\item \textbf{\(\ell_1\)-CS:} scale-normalized complex 3D discrete cosine transform (DCT) recovery of \(f\) and \(g\), using phase-preserving complex soft thresholding, \(\rho=20\), at most \(800\) iterations, and tolerance \(10^{-6}\).
\item \textbf{Standard joint 3D ALOHA:} joint three-channel Hankel completion without Maxwell projection.
\item \textbf{Proposed MC-Hankel:} Algorithm~\ref{alg:mc_hankel}.
\end{enumerate}
Both structured-Hankel completions use a \(3\times3\times3\) patch, \(\mu=10\), and \(30\) iterations. The target rank is \(r=22\) for J1 and \(r=30\) for J2 source models. These source-specific settings are fixed across all sampling rates, SNRs, and trials. The proposed and standard methods, therefore, differ in the Maxwell projection, noise fidelity, and conjugate symmetry but not in the low-rank budget.

\subsection{Evaluation metrics and statistical analysis}
For reference magnitude volume \(V\) and reconstruction \(\widehat{V}\), the full-volume errors are the normalized-mean squared error \(\mathrm{NMSE}_{3D}\) and relative $\ell_2$ error ($\mathrm{Rel.}L_2$),  
\begin{align}
\mathrm{NMSE}_{3D}
&=\frac{\norm{\widehat V-V}_2^2}{\norm{V}_2^2},\\
\mathrm{Rel.}L_2
&=\frac{\norm{\widehat V-V}_2}{\norm{V}_2}.
\end{align}
Support topology is measured by Dice overlap after thresholding each volume at \(25\%\) and \(55\%\) of the reference peak. 
The model residual used to assess physical compatibility is
\begin{equation}
r_{\Model}(\tJ)
=\dfrac{\norm{\tJ-\Proj_{\Model}(\tJ)}_F}
{\max\left(\norm{\tJ}_F,\epsilon\right)},
\label{eq:model_residual}
\end{equation}
Structural similarity is calculated on non-negligible slices in all three orientations and averaged to obtain tri-planar SSIM, and PSNR is evaluated using the reference dynamic range. The runtimes are wall-clock measurements from the same experiment run.

Since all methods share each random realization, the comparisons are paired. For each source-noise-rate configuration, the analysis reports paired mean difference, an exact two-sided Wilcoxon signed-rank test, Holm-adjusted \(p\)-values within each comparator-metric family, a \(10,000\)-resample paired bootstrap confidence interval, paired Cohen's \(d_z\), rank-biserial effect size, and win rate. With only five pairs, the smallest attainable two-sided exact Wilcoxon \(p\)-value is \(0.0625\); therefore, the inferential analysis is interpreted as preliminary, and effect sizes and bootstrap intervals are emphasized.

\section{Numerical simulations and results} \label{s:results}

In this section, we evaluated MC-Hankel through comprehensive qualitative and quantitative experiments, comparing its performance against the considered baseline approaches. 

\subsection{2D qualitative reconstruction}
\begin{figure*}[!htb]
\centering
\includegraphics[width=\textwidth]{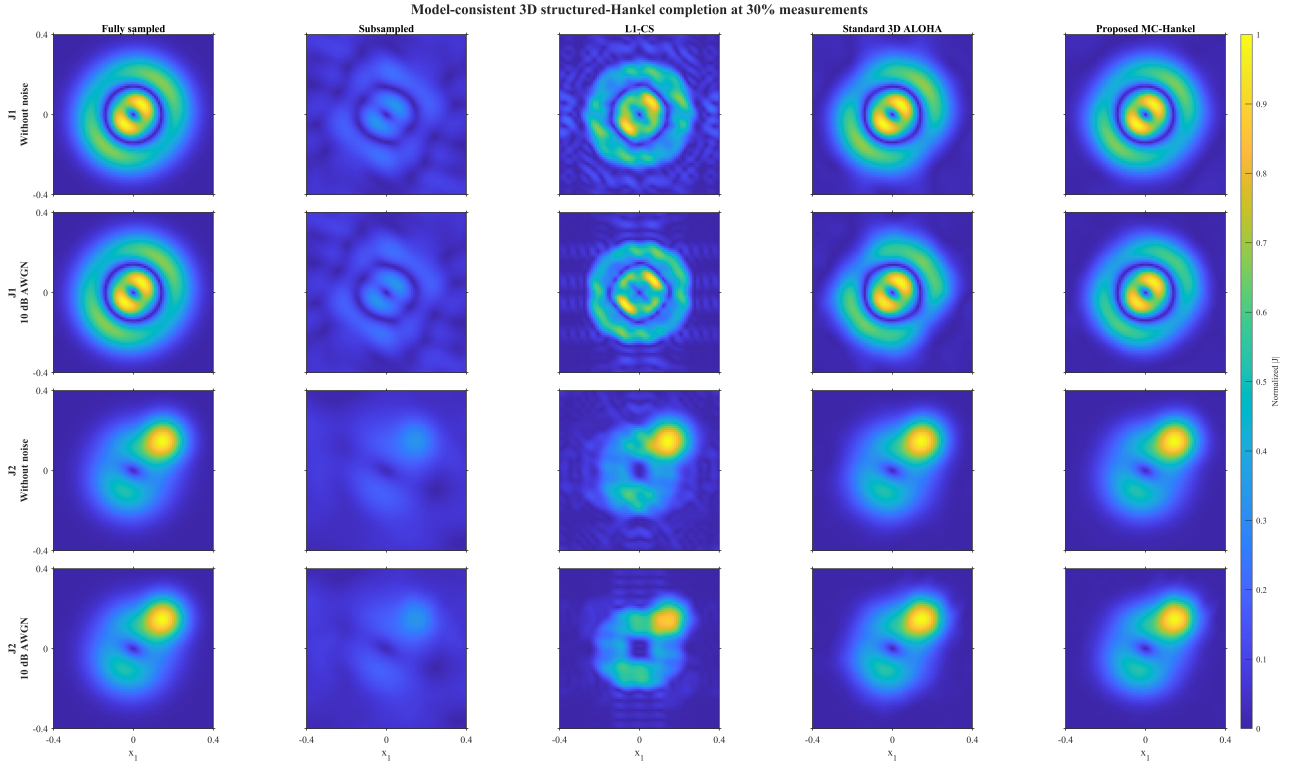}
\caption{Central-slice comparison at \(30\%\) measurements. Rows correspond to J1 and J2 under clean and \(10\) dB AWGN conditions. Columns show the fully sampled reference, zero-filled sub-sampling, \(\ell_1\)-CS, standard joint 3D ALOHA, and the proposed MC-Hankel method. Display normalization is shared within each row.}
\label{fig:qualitative}
\end{figure*}

Fig.~\ref{fig:qualitative} shows central \(x_3=0\) slices at the most challenging \(30\%\) measurement rate. Zero-filling Fourier inversion retains strong under-sampling artefacts, and the DCT \(\ell_1\) baseline only partially restores the source. Although standard 3D ALOHA recovers the dominant structures, MC-Hankel suppresses residual out-of-model structures and improves continuity of both the diffused J1 source and the mixed J2 source. These observations persist under \(10\) dB AWGN.

\subsection{Full-volume accuracy}
Tables~\ref{tab:mc_j1} and \ref{tab:mc_j2} report every sampling rate and noise condition. MC-Hankel achieves the best mean PSNR and SSIM in all twelve configurations. Relative to standard ALOHA, the PSNR gain ranges from \(0.653\) dB for noisy J1 at \(40\%\) measurements to \(4.217\) dB for clean J2 at \(30\%\). The corresponding SSIM gain ranges from \(0.0266\) to \(0.0703\). For J1, the largest clean-data gain appears at \(50\%\), where PSNR increases from \(38.87\) to \(42.50\) dB and SSIM from \(0.4865\) to \(0.5568\). For J2, the proposed method reaches \(52.70\) to \(54.89\) dB in the clean experiments and remains superior under \(10\) dB AWGN.

\begin{table*}[!htb]
\centering
\small
\caption{Full-volume reconstruction for J1: curl-type smooth source  $\widehat{\mathbf p}\times\nabla g$ over five independent paired Monte Carlo trials. PSNR and tri-planar SSIM are reported as mean $\pm$ standard deviation. The last column reports the relative Maxwell-model residual.}
\label{tab:mc_j1}
\begin{tabular}{cclccc}
\toprule
Scenario & Meas. (\%) & Method & PSNR & SSIM & Model res. \\
\midrule

No noise 
& 30 & Sub. 
& 19.71 $\pm$ 0.10 
& 0.1260 $\pm$ 0.0052 
& 1.45e-16 \\
& & $\ell_1$-CS 
& 23.06 $\pm$ 2.17 
& 0.1650 $\pm$ 0.0627 
& 1.49e-16 \\
& & Std. ALOHA 
& 33.54 $\pm$ 1.46 
& 0.4313 $\pm$ 0.0193 
& 1.15e-01 \\
& & \textbf{Proposed} 
& \textbf{36.10 $\pm$ 1.41} 
& \textbf{0.4799 $\pm$ 0.0156} 
& 1.47e-16 \\
\cmidrule(lr){2-6}

& 40 & Sub. 
& 20.35 $\pm$ 0.14 
& 0.1397 $\pm$ 0.0065 
& 1.46e-16 \\
& & $\ell_1$-CS 
& 24.67 $\pm$ 2.24 
& 0.1957 $\pm$ 0.0378 
& 1.53e-16 \\
& & Std. ALOHA 
& 36.32 $\pm$ 1.53 
& 0.4604 $\pm$ 0.0243 
& 7.66e-02 \\
& & \textbf{Proposed} 
& \textbf{38.80 $\pm$ 1.18} 
& \textbf{0.5122 $\pm$ 0.0100} 
& 1.35e-16 \\
\cmidrule(lr){2-6}

& 50 & Sub. 
& 21.20 $\pm$ 0.24 
& 0.1600 $\pm$ 0.0120 
& 1.40e-16 \\
& & $\ell_1$-CS 
& 27.49 $\pm$ 2.88 
& 0.2519 $\pm$ 0.0635 
& 1.57e-16 \\
& & Std. ALOHA 
& 38.87 $\pm$ 1.37 
& 0.4865 $\pm$ 0.0178 
& 6.56e-02 \\
& & \textbf{Proposed} 
& \textbf{42.50 $\pm$ 1.45} 
& \textbf{0.5568 $\pm$ 0.0228} 
& 1.30e-16 \\

\midrule

10 dB AWGN 
& 30 & Sub. 
& 19.70 $\pm$ 0.10 
& 0.1247 $\pm$ 0.0049 
& 1.14e-01 \\
& & $\ell_1$-CS 
& 22.71 $\pm$ 0.51 
& 0.1939 $\pm$ 0.0470 
& 1.51e-16 \\
& & Std. ALOHA 
& 29.45 $\pm$ 0.86 
& 0.4718 $\pm$ 0.0200 
& 1.54e-01 \\
& & \textbf{Proposed} 
& \textbf{30.77 $\pm$ 0.74} 
& \textbf{0.5068 $\pm$ 0.0130} 
& 1.58e-16 \\
\cmidrule(lr){2-6}

& 40 & Sub. 
& 20.33 $\pm$ 0.15 
& 0.1380 $\pm$ 0.0064 
& 1.19e-01 \\
& & $\ell_1$-CS 
& 24.25 $\pm$ 0.14 
& 0.2315 $\pm$ 0.0810 
& 1.54e-16 \\
& & Std. ALOHA 
& 32.12 $\pm$ 0.89 
& 0.5110 $\pm$ 0.0159 
& 1.01e-01 \\
& & \textbf{Proposed} 
& \textbf{32.77 $\pm$ 0.63} 
& \textbf{0.5377 $\pm$ 0.0134} 
& 1.51e-16 \\
\cmidrule(lr){2-6}

& 50 & Sub. 
& 21.16 $\pm$ 0.24 
& 0.1582 $\pm$ 0.0120 
& 1.17e-01 \\
& & $\ell_1$-CS 
& 26.03 $\pm$ 0.79 
& 0.2846 $\pm$ 0.0462 
& 1.57e-16 \\
& & Std. ALOHA 
& 34.37 $\pm$ 0.78 
& 0.5248 $\pm$ 0.0073 
& 7.54e-02 \\
& & \textbf{Proposed} 
& \textbf{35.05 $\pm$ 0.83} 
& \textbf{0.5587 $\pm$ 0.0137} 
& 1.57e-16 \\

\bottomrule
\end{tabular}
\end{table*}

\begin{table*}[!htb]
\centering
\small
\caption{Full-volume reconstruction for J2: mixed source $f\widehat{\mathbf p}+\widehat{\mathbf p}\times\nabla g$ over five independent paired Monte Carlo trials. PSNR and tri-planar SSIM are reported as mean $\pm$ standard deviation. The last column reports the relative Maxwell-model residual.}
\label{tab:mc_j2}
\begin{tabular}{cclccc}
\toprule
Scenario & Meas. (\%) & Method & PSNR & SSIM & Model res. \\
\midrule

No noise 
& 30 & Sub. 
& 23.06 $\pm$ 0.31 
& 0.1332 $\pm$ 0.0110 
& 1.20e-16 \\
& & $\ell_1$-CS 
& 27.83 $\pm$ 1.74 
& 0.1864 $\pm$ 0.0389 
& 1.50e-16 \\
& & Std. ALOHA 
& 48.48 $\pm$ 1.51 
& 0.5731 $\pm$ 0.0110 
& 3.12e-02 \\
& & \textbf{Proposed} 
& \textbf{52.70 $\pm$ 0.99} 
& \textbf{0.6250 $\pm$ 0.0067} 
& 1.45e-16 \\
\cmidrule(lr){2-6}

& 40 & Sub. 
& 23.51 $\pm$ 0.34 
& 0.1395 $\pm$ 0.0109 
& 1.22e-16 \\
& & $\ell_1$-CS 
& 30.60 $\pm$ 1.34 
& 0.2111 $\pm$ 0.0270 
& 1.52e-16 \\
& & Std. ALOHA 
& 51.06 $\pm$ 1.00 
& 0.5909 $\pm$ 0.0083 
& 2.39e-02 \\
& & \textbf{Proposed} 
& \textbf{53.85 $\pm$ 0.68} 
& \textbf{0.6223 $\pm$ 0.0083} 
& 1.48e-16 \\
\cmidrule(lr){2-6}

& 50 & Sub. 
& 24.16 $\pm$ 0.41 
& 0.1513 $\pm$ 0.0141 
& 1.30e-16 \\
& & $\ell_1$-CS 
& 31.75 $\pm$ 2.57 
& 0.2389 $\pm$ 0.0478 
& 1.52e-16 \\
& & Std. ALOHA 
& 52.62 $\pm$ 0.62 
& 0.6021 $\pm$ 0.0140 
& 2.12e-02 \\
& & \textbf{Proposed} 
& \textbf{54.89 $\pm$ 0.61} 
& \textbf{0.6367 $\pm$ 0.0070} 
& 1.42e-16 \\

\midrule

10 dB AWGN 
& 30 & Sub. 
& 23.02 $\pm$ 0.29 
& 0.1290 $\pm$ 0.0113 
& 1.26e-16 \\
& & $\ell_1$-CS 
& 28.63 $\pm$ 1.46 
& 0.2697 $\pm$ 0.0310 
& 1.65e-16 \\
& & Std. ALOHA 
& 40.60 $\pm$ 2.32 
& 0.5081 $\pm$ 0.0971 
& 5.47e-02 \\
& & \textbf{Proposed} 
& \textbf{42.66 $\pm$ 2.41} 
& \textbf{0.5506 $\pm$ 0.0994} 
& 1.63e-16 \\
\cmidrule(lr){2-6}

& 40 & Sub. 
& 23.44 $\pm$ 0.38 
& 0.1322 $\pm$ 0.0149 
& 1.25e-16 \\
& & $\ell_1$-CS 
& 30.98 $\pm$ 1.26 
& 0.3034 $\pm$ 0.0754 
& 1.54e-16 \\
& & Std. ALOHA 
& 42.11 $\pm$ 3.75 
& 0.4757 $\pm$ 0.1138 
& 4.24e-02 \\
& & \textbf{Proposed} 
& \textbf{43.33 $\pm$ 4.03} 
& \textbf{0.5064 $\pm$ 0.1178} 
& 1.61e-16 \\
\cmidrule(lr){2-6}

& 50 & Sub. 
& 24.06 $\pm$ 0.45 
& 0.1440 $\pm$ 0.0157 
& 1.24e-16 \\
& & $\ell_1$-CS 
& 32.23 $\pm$ 0.86 
& 0.3405 $\pm$ 0.0768 
& 1.62e-16 \\
& & Std. ALOHA 
& 42.90 $\pm$ 3.34 
& 0.4644 $\pm$ 0.0919 
& 3.78e-02 \\
& & \textbf{Proposed} 
& \textbf{43.85 $\pm$ 3.58} 
& \textbf{0.4924 $\pm$ 0.0942} 
& 1.57e-16 \\

\bottomrule
\end{tabular}
\end{table*}

\begin{figure*}[!htb]
\centering
\includegraphics[width=\textwidth]{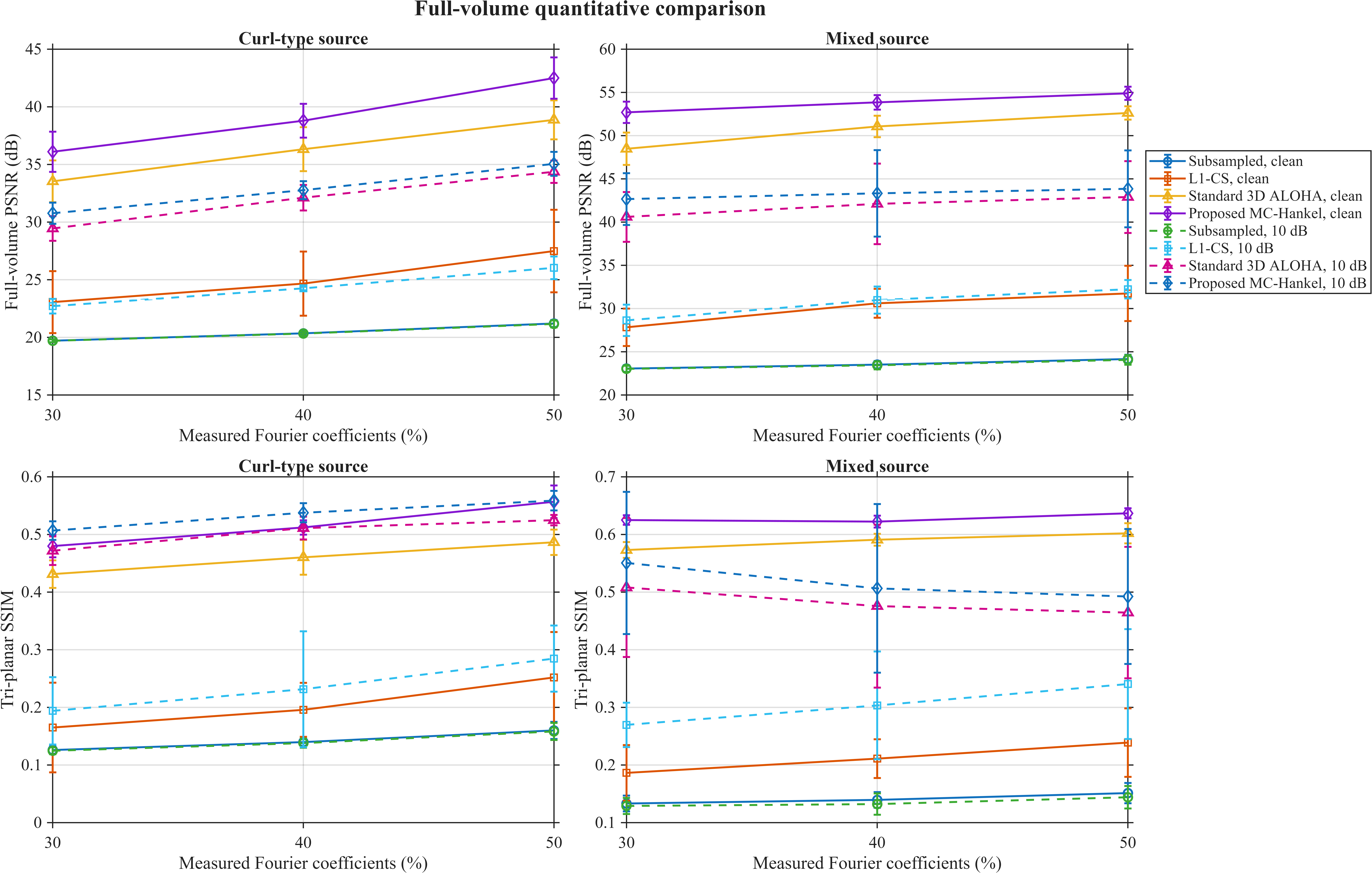}
\caption{Full-volume PSNR and tri-planar SSIM versus measurement rate. Solid lines show clean data and dashed lines show \(10\) dB AWGN. Error bars are \(95\%\) Student-\(t\) confidence intervals across five paired trials.}
\label{fig:metrics}
\end{figure*}

Fig.~\ref{fig:metrics} summarizes the same full-volume trends with confidence intervals. Increasing the sampling rate improves all methods, but the gap between the zero-filled and structured-Hankel reconstructions remains substantial. The proposed method consistently performs better than the standard ALOHA, and the gain is particularly pronounced in clean low-sampling J2 and clean high-sampling J1.

\subsection{Maxwell-model consistency}
\begin{figure*}[!htb]
\centering
\includegraphics[width=\textwidth]{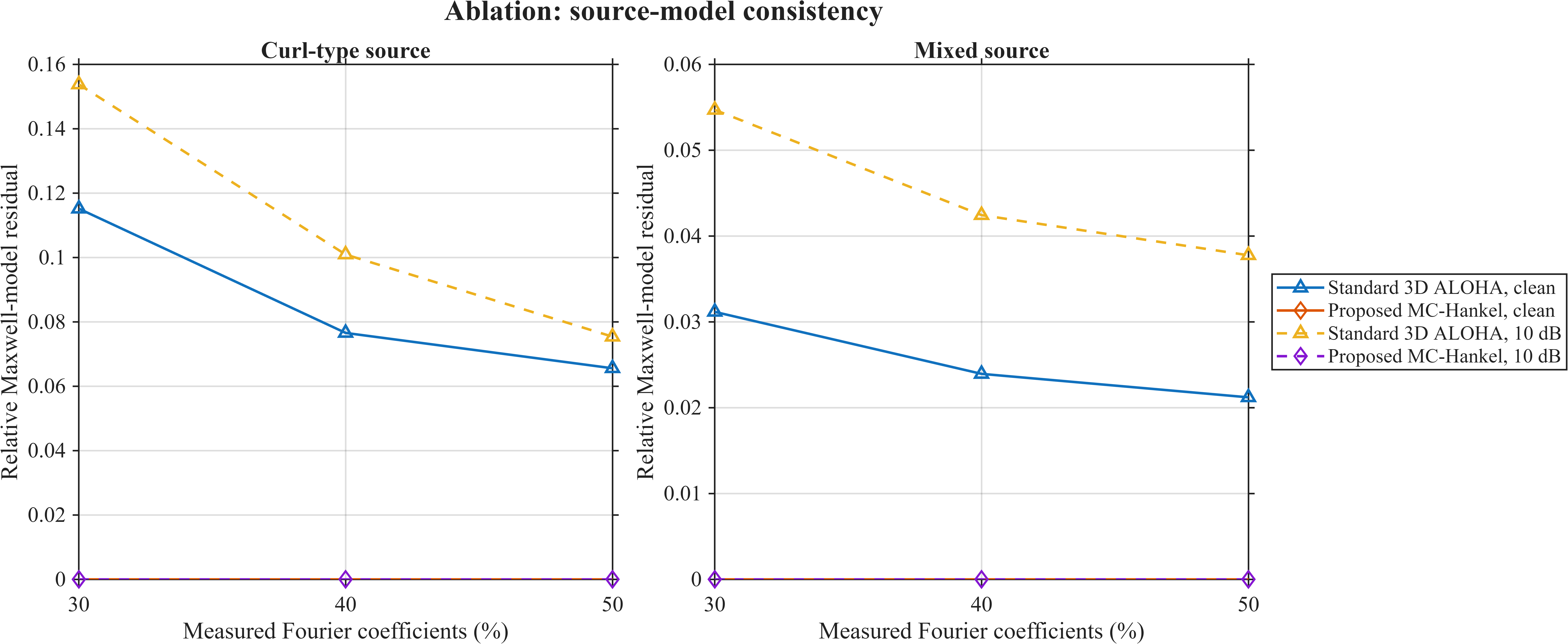}
\caption{Relative Maxwell-model residual for standard joint 3D ALOHA and the proposed MC-Hankel method. The proposed reconstruction remains in the admissible source subspace to numerical precision.}
\label{fig:model_residual}
\end{figure*}

The proposed projection removes the component orthogonal to the admissible source subspace $\Model$. Fig.~\ref{fig:model_residual} shows that standard ALOHA has mean model residuals between \(2.12\times10^{-2}\) and \(1.54\times10^{-1}\), whereas MC-Hankel reduces the residual to approximately \(10^{-16}\) in every configuration. The model projection is not a post-hoc patch. Instead, it is applied within each iteration to ensure that the low-rank and physical consistency are optimized concurrently.

\subsection{3D topology and error localization}
Table~\ref{tab:topology_metrics} evaluates the complete volumes. MC-Hankel reduces Rel.\(L_2\) in every configuration. The absolute reduction ranges from \(0.0059\) to \(0.0427\), equivalent to a relative reduction of \(7.4\%\)--\(38.9\%\) compared to standard ALOHA. Dice$_{25}$ also improves consistently, indicating better recovery of the overall source support. Dice$_{55}$ is further mixed for noisy J1 because exact model projection can slightly attenuate or reshape the strongest core even while reducing global error; it illustrates why topology should be assessed at more than one threshold.

\begin{table*}[!htb]
\centering\scriptsize
\caption{Full-volume relative error and support-overlap metrics. Values are mean $\pm$ standard deviation over five paired trials. Lower Rel.$L_2$ is better; higher Dice is better.}
\label{tab:topology_metrics}
\resizebox{\textwidth}{!}{%
\begin{tabular}{llc ccc ccc}
\toprule
&&\multicolumn{3}{c}{Standard 3D ALOHA}&\multicolumn{3}{c}{Proposed MC-Hankel}\\
Source & Scenario & Meas. (\%) & Rel.$L_2$ & Dice$_{25}$ & Dice$_{55}$ & Rel.$L_2$ & Dice$_{25}$ & Dice$_{55}$\\
\midrule
J1 & Clean & 30 & 0.1670 $\pm$ 0.0280 & 0.9145 $\pm$ 0.0213 & 0.8647 $\pm$ 0.0284 & \textbf{0.1243 $\pm$ 0.0214} & \textbf{0.9365 $\pm$ 0.0144} & 0.8913 $\pm$ 0.0270 \\
 &  & 40 & 0.1213 $\pm$ 0.0218 & 0.9442 $\pm$ 0.0166 & 0.9031 $\pm$ 0.0400 & \textbf{0.0908 $\pm$ 0.0124} & \textbf{0.9605 $\pm$ 0.0113} & 0.9246 $\pm$ 0.0361 \\
 &  & 50 & 0.0903 $\pm$ 0.0150 & 0.9659 $\pm$ 0.0100 & 0.9407 $\pm$ 0.0265 & \textbf{0.0595 $\pm$ 0.0101} & \textbf{0.9787 $\pm$ 0.0058} & 0.9545 $\pm$ 0.0165 \\
\addlinespace
J1 & 10 dB AWGN & 30 & 0.2655 $\pm$ 0.0263 & 0.8221 $\pm$ 0.0204 & 0.7608 $\pm$ 0.0246 & \textbf{0.2277 $\pm$ 0.0193} & \textbf{0.8506 $\pm$ 0.0141} & 0.7181 $\pm$ 0.0451 \\
 &  & 40 & 0.1953 $\pm$ 0.0198 & 0.8754 $\pm$ 0.0149 & 0.8335 $\pm$ 0.0483 & \textbf{0.1807 $\pm$ 0.0131} & \textbf{0.8877 $\pm$ 0.0122} & 0.7993 $\pm$ 0.0636 \\
 &  & 50 & 0.1506 $\pm$ 0.0140 & 0.9097 $\pm$ 0.0123 & 0.8711 $\pm$ 0.0463 & \textbf{0.1392 $\pm$ 0.0129} & \textbf{0.9165 $\pm$ 0.0097} & 0.8485 $\pm$ 0.0348 \\
\addlinespace
\midrule
J2 & Clean & 30 & 0.0417 $\pm$ 0.0077 & 0.9880 $\pm$ 0.0048 & 0.9904 $\pm$ 0.0036 & \textbf{0.0255 $\pm$ 0.0029} & \textbf{0.9933 $\pm$ 0.0016} & 0.9893 $\pm$ 0.0044 \\
 &  & 40 & 0.0308 $\pm$ 0.0037 & 0.9926 $\pm$ 0.0013 & 0.9930 $\pm$ 0.0035 & \textbf{0.0223 $\pm$ 0.0018} & \textbf{0.9952 $\pm$ 0.0011} & 0.9933 $\pm$ 0.0017 \\
 &  & 50 & 0.0256 $\pm$ 0.0018 & 0.9949 $\pm$ 0.0010 & 0.9939 $\pm$ 0.0026 & \textbf{0.0197 $\pm$ 0.0014} & \textbf{0.9957 $\pm$ 0.0008} & 0.9934 $\pm$ 0.0032 \\
\addlinespace
J2 & 10 dB AWGN & 30 & 0.1052 $\pm$ 0.0298 & 0.9620 $\pm$ 0.0120 & 0.9424 $\pm$ 0.0336 & \textbf{0.0833 $\pm$ 0.0264} & \textbf{0.9730 $\pm$ 0.0093} & 0.9413 $\pm$ 0.0240 \\
 &  & 40 & 0.0924 $\pm$ 0.0395 & 0.9680 $\pm$ 0.0166 & 0.9479 $\pm$ 0.0302 & \textbf{0.0813 $\pm$ 0.0379} & \textbf{0.9740 $\pm$ 0.0147} & 0.9476 $\pm$ 0.0272 \\
 &  & 50 & 0.0831 $\pm$ 0.0318 & 0.9740 $\pm$ 0.0116 & 0.9521 $\pm$ 0.0255 & \textbf{0.0751 $\pm$ 0.0305} & \textbf{0.9779 $\pm$ 0.0106} & 0.9543 $\pm$ 0.0232 \\
\addlinespace
\bottomrule
\end{tabular}}
\end{table*}

\begin{figure*}[!htb]
\centering
\includegraphics[width=\textwidth]{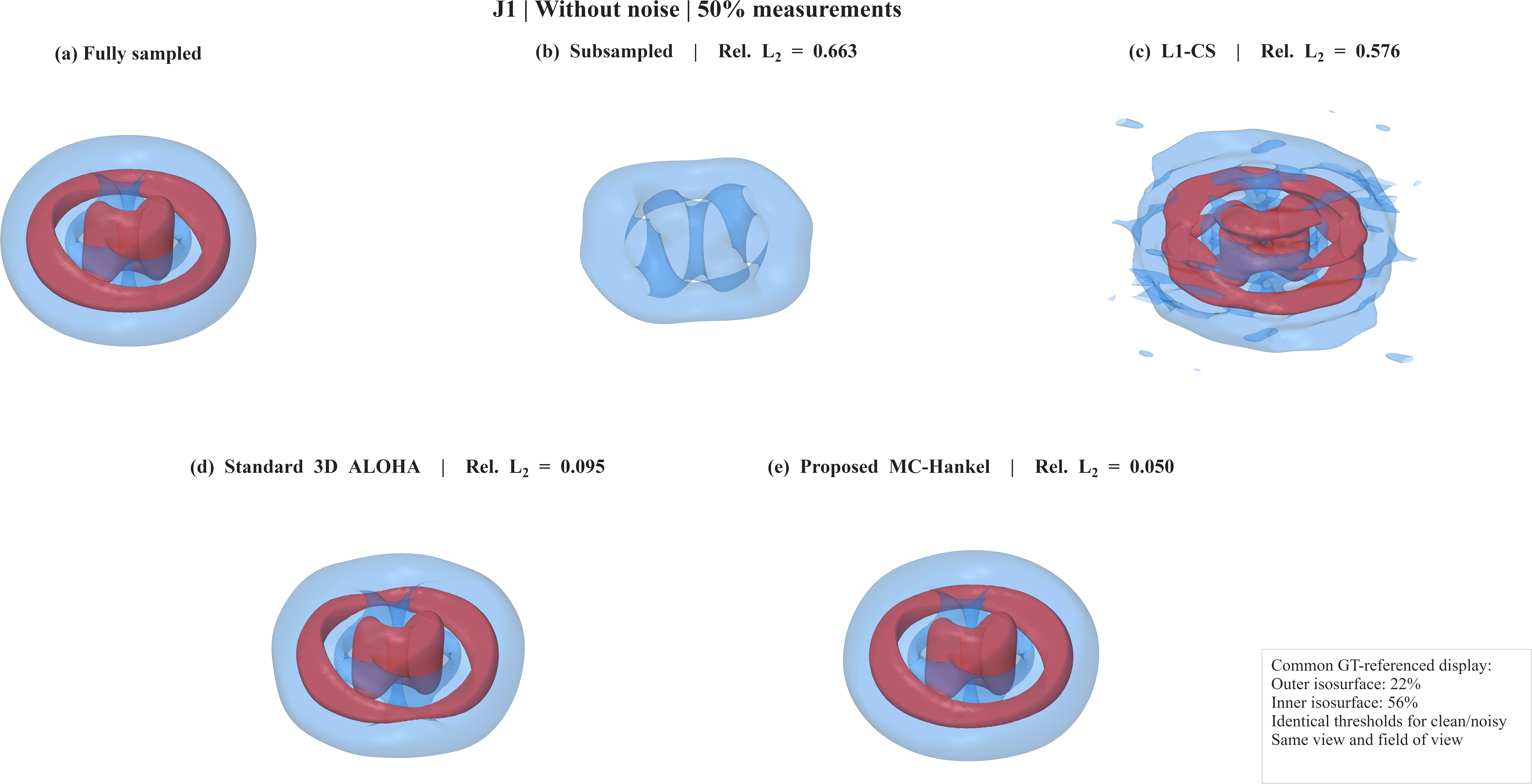}
\caption{J1 full-volume reconstruction at \(50\%\) measurements without noise. Common axes and absolute isosurface thresholds are used for all methods.}
\label{fig:j1clean}
\end{figure*}

\begin{figure*}[!htb]
\centering
\includegraphics[width=\textwidth]{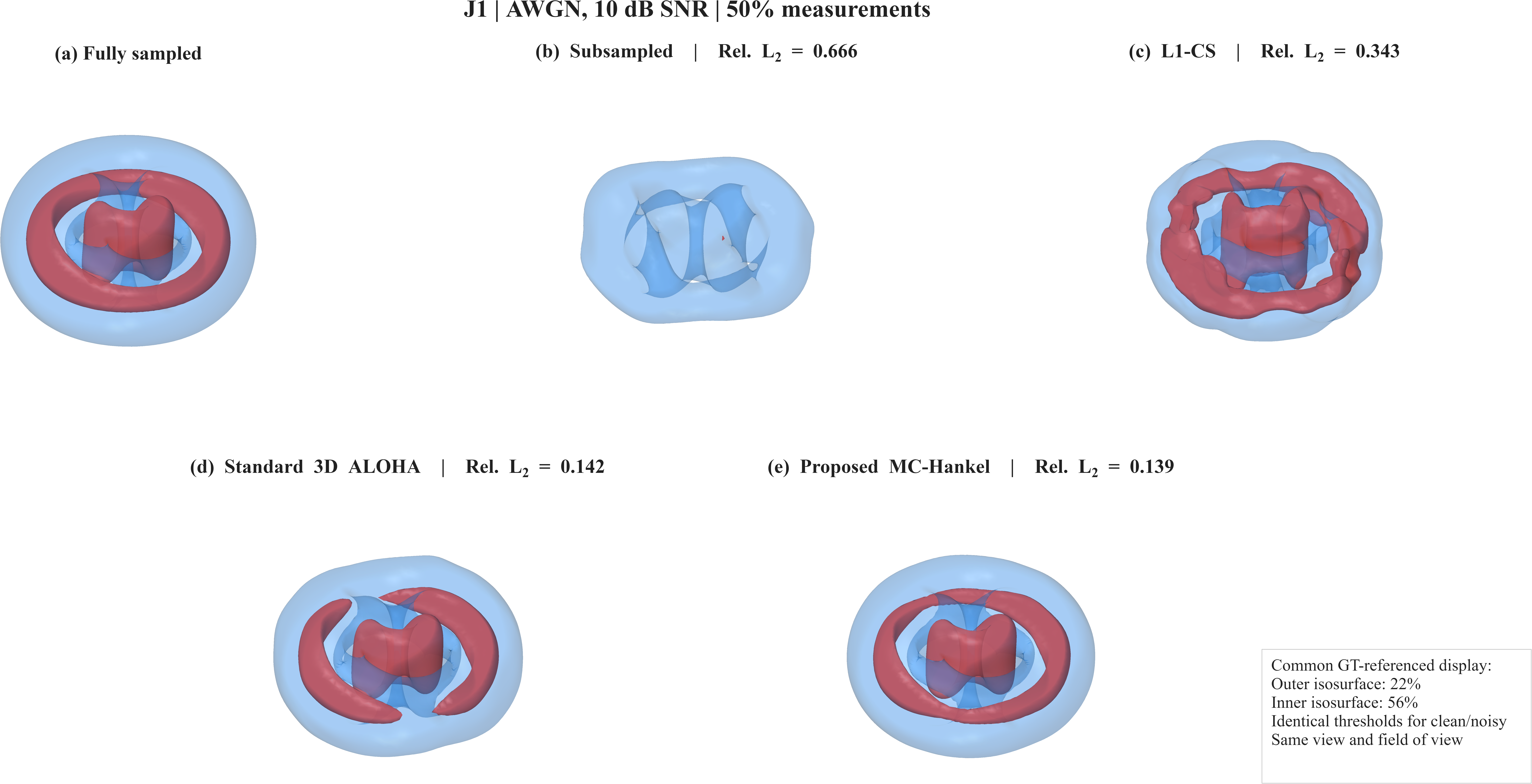}
\caption{J1 full-volume reconstruction at \(50\%\) measurements with \(10\) dB AWGN.}
\label{fig:j1noisy}
\end{figure*}

\begin{figure*}[!htb]
\centering
\includegraphics[width=\textwidth]{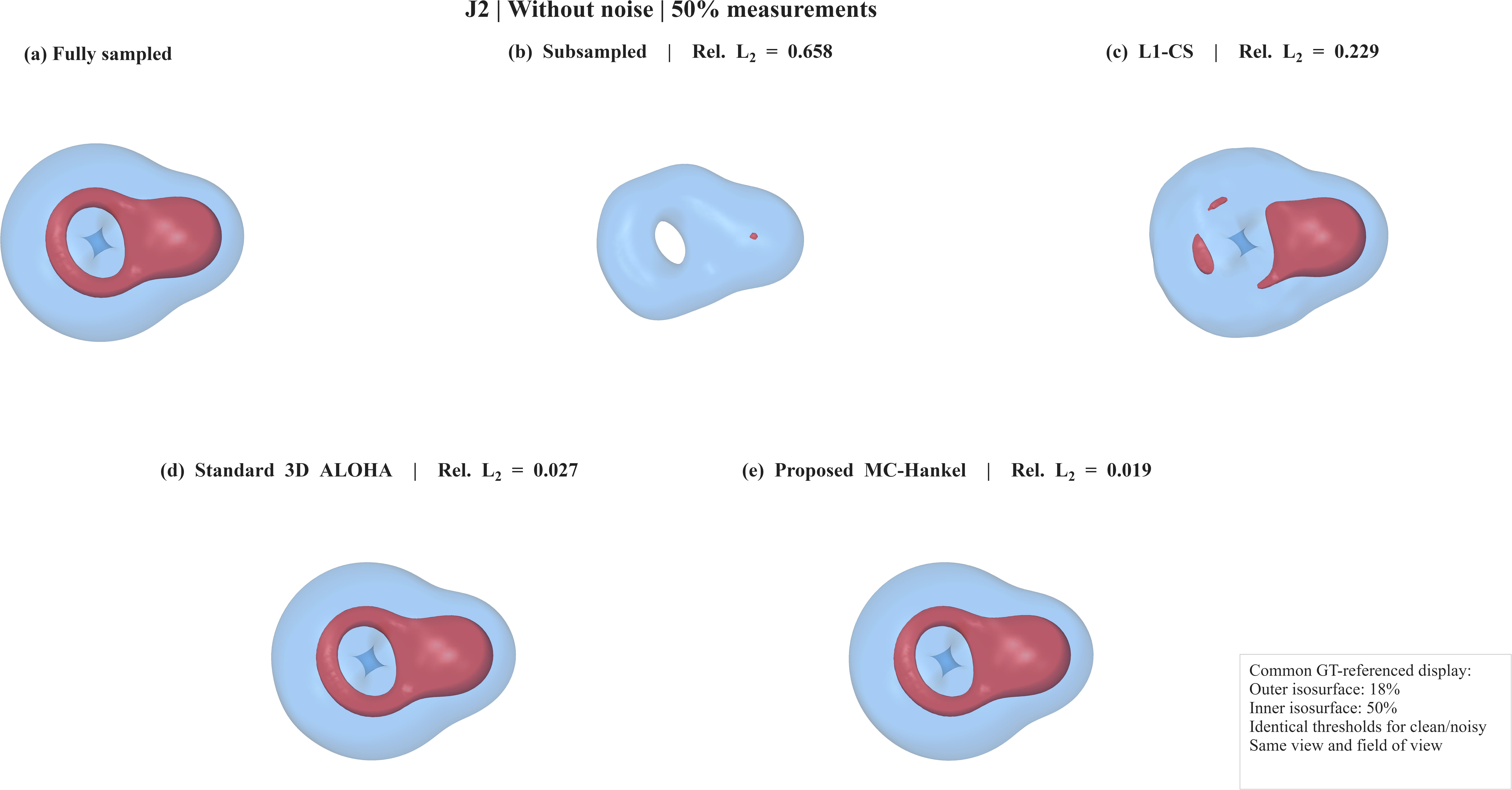}
\caption{J2 full-volume reconstruction at \(50\%\) measurements without noise.}
\label{fig:j2clean}
\end{figure*}

\begin{figure*}[!htb]
\centering
\includegraphics[width=\textwidth]{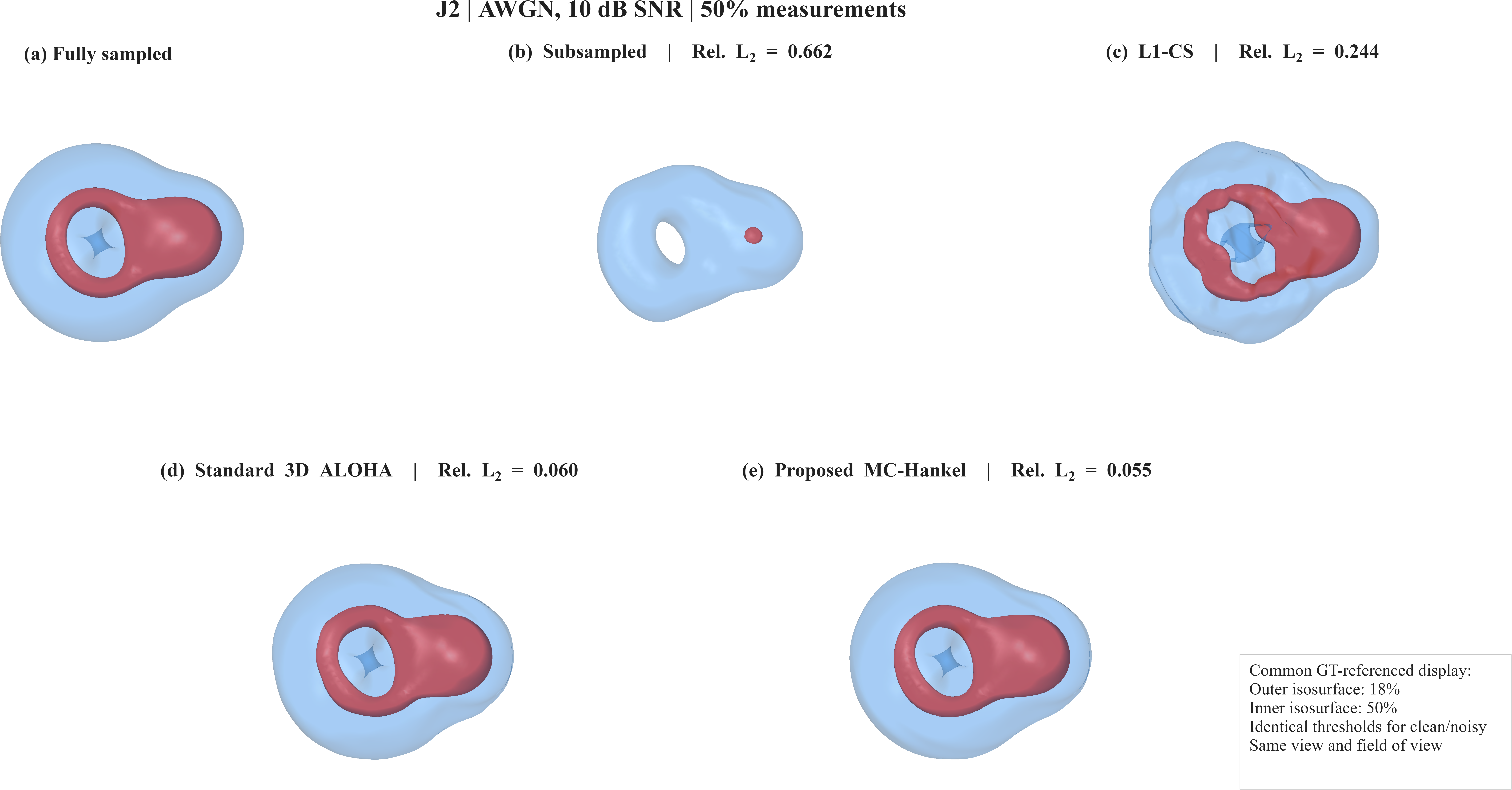}
\caption{J2 full-volume reconstruction at \(50\%\) measurements with \(10\) dB AWGN.}
\label{fig:j2noisy}
\end{figure*}

Figs.~\ref{fig:j1clean}-\ref{fig:j2noisy} show representative trial-1 reconstructions at \(50\%\) measurements. All panels use common physical axes and isosurface thresholds at \(20\%\) and \(55\%\) of the reference maximum. This shared scaling prevents visual improvement from being created by independent normalization. For J1, MC-Hankel better preserves the diffused outer support and the compact high-intensity component, particularly under noise. For J2, both Hankel methods recover the main source well, but the MC-Hankel result more closely matches the reference amplitude and surface geometry.

\begin{figure*}[!htb]
\centering
\includegraphics[height=0.72\textheight,keepaspectratio]{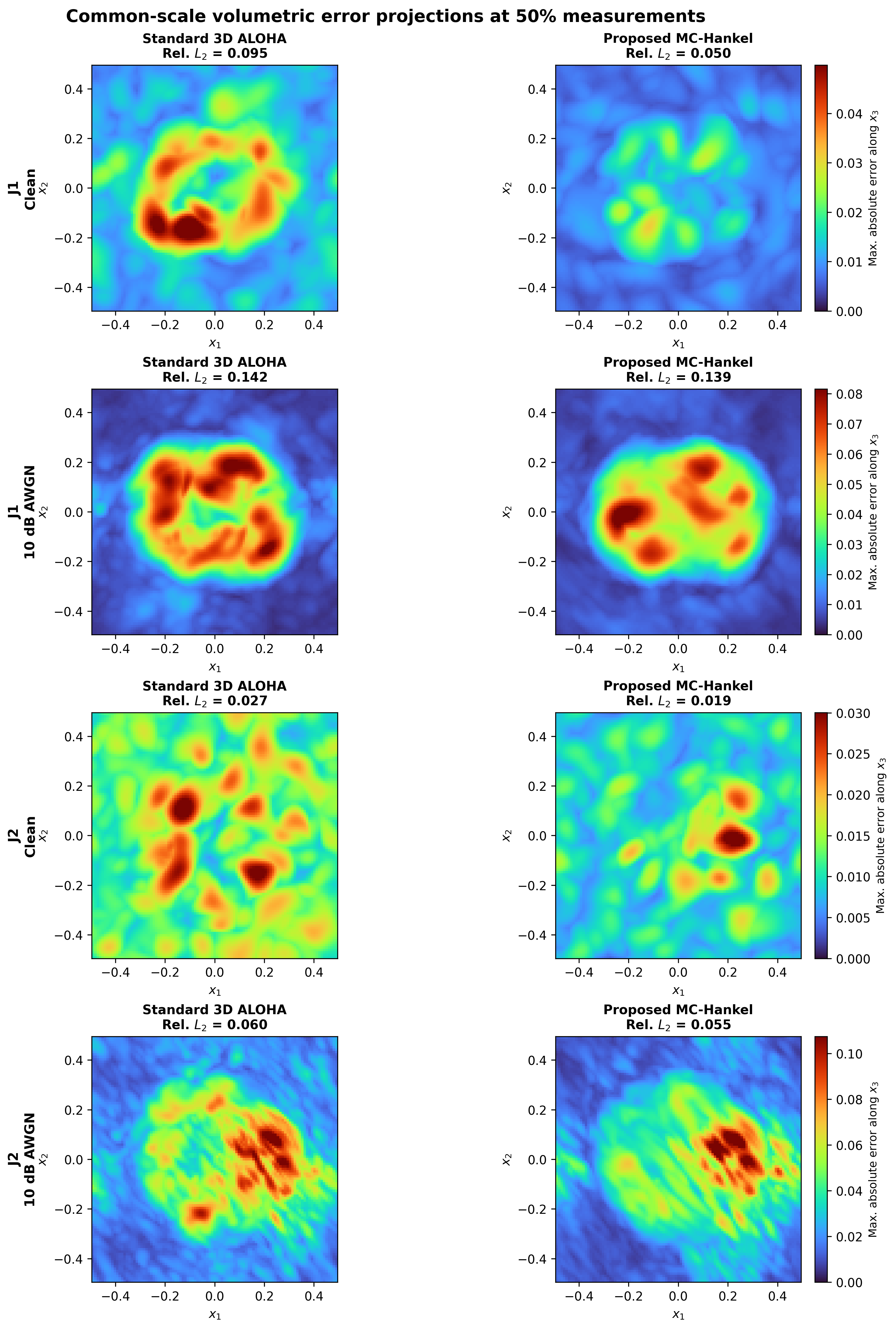}
\caption{Common-scale maximum absolute error projections along \(x_3\) at \(50\%\) measurements. Each row compares standard 3D ALOHA (left) with MC-Hankel (right) for J1 clean, J1 noisy, J2 clean, and J2 noisy conditions. \(\mathrm{Rel.}L_2\) values are computed from the original \(101\times 101\times 101\) volumes before display interpolation.}
\label{fig:error_projection}
\end{figure*}

The maximum-intensity projections of absolute volume error in Fig.~\ref{fig:error_projection} localize the remaining discrepancies. For all four representative cases, MC-Hankel produces a lower relative \(L_2\) error than standard ALOHA. Error reduction is concentrated around source boundaries and high-amplitude cores, consistent with the improvements in PSNR, SSIM, Rel.\(L_2\), and Dice$_{25}$.

\subsection{Paired statistical analysis}
Table~\ref{tab:paired_statistics} reports the paired advantage over standard ALOHA. The proposed method wins every PSNR and SSIM pair in all twelve configurations, and all paired bootstrap intervals are positive. Cohen's \(d_z\) ranges from \(1.21\) to \(4.89\) for PSNR and from \(2.22\) to \(7.58\) for SSIM, indicating large standardized paired effects in this experiment. Nevertheless, the exact two-sided signed-rank \(p\)-value is \(0.0625\) for a complete five-of-five win and becomes \(0.75\) after Holm correction across twelve settings. Accordingly, these results demonstrate consistent paired improvements but do not justify a formal claim of statistical significance at \(\alpha=0.05\). A larger confirmatory Monte Carlo run is required.

\begin{table*}[!htb]
\centering\scriptsize
\caption{Paired improvements of the proposed method over standard 3D ALOHA. Bootstrap intervals use 10,000 paired resamples. With only five pairs, the smallest attainable two-sided exact Wilcoxon $p$-value is 0.0625; consequently, no comparison survives Holm correction despite a 100\% win rate.}
\label{tab:paired_statistics}
\resizebox{\textwidth}{!}{%
\begin{tabular}{llc ccc ccc}
\toprule
&&&\multicolumn{3}{c}{PSNR improvement (dB)}&\multicolumn{3}{c}{SSIM improvement}\\
Source&Scenario&Meas. (\%)&Mean $\Delta$&Bootstrap 95\% CI&$d_z$&Mean $\Delta$&Bootstrap 95\% CI&$d_z$\\
\midrule
J1 & Clean & 30 & 2.564 & [1.619, 3.437] & 2.19 & 0.0486 & [0.0399, 0.0583] & 4.16 \\
J1 & Clean & 40 & 2.471 & [1.692, 3.171] & 2.58 & 0.0518 & [0.0375, 0.0648] & 3.04 \\
J1 & Clean & 50 & 3.627 & [2.598, 4.762] & 2.69 & 0.0703 & [0.0541, 0.0868] & 3.54 \\
J1 & 10 dB AWGN & 30 & 1.323 & [0.759, 1.707] & 2.13 & 0.0350 & [0.0291, 0.0459] & 2.89 \\
J1 & 10 dB AWGN & 40 & 0.653 & [0.265, 1.109] & 1.21 & 0.0266 & [0.0197, 0.0345] & 2.79 \\
J1 & 10 dB AWGN & 50 & 0.685 & [0.315, 1.094] & 1.38 & 0.0339 & [0.0209, 0.0449] & 2.22 \\
J2 & Clean & 30 & 4.217 & [3.287, 5.131] & 3.55 & 0.0519 & [0.0466, 0.0570] & 7.58 \\
J2 & Clean & 40 & 2.790 & [2.007, 3.507] & 2.88 & 0.0314 & [0.0214, 0.0393] & 2.79 \\
J2 & Clean & 50 & 2.271 & [1.933, 2.657] & 4.89 & 0.0347 & [0.0243, 0.0454] & 2.44 \\
J2 & 10 dB AWGN & 30 & 2.059 & [1.531, 2.850] & 2.35 & 0.0425 & [0.0368, 0.0497] & 5.11 \\
J2 & 10 dB AWGN & 40 & 1.218 & [0.914, 1.542] & 3.03 & 0.0307 & [0.0268, 0.0346] & 6.05 \\
J2 & 10 dB AWGN & 50 & 0.951 & [0.575, 1.269] & 2.08 & 0.0279 & [0.0236, 0.0315] & 5.45 \\
\bottomrule
\end{tabular}}
\end{table*}

\subsection{Computational cost}
Table~\ref{tab:runtime} shows that the proposed projection adds modest overhead to standard ALOHA: mean time increases from \(0.989\) to \(1.160\) s, approximately \(17.2\%\). Its mean runtime is essentially the same as the corrected \(\ell_1\)-CS baseline. These values should be interpreted comparatively because hardware metadata were not retained with the run.

\begin{table}[!htb]
\centering
\caption{Wall-clock reconstruction times averaged across sources, sampling rates, and noise conditions. These are comparative measurements from the same MATLAB experiment run; hardware metadata were not retained.}
\label{tab:runtime}
\begin{tabular}{lccc}
\toprule
Method & Mean (s) & Minimum (s) & Maximum (s)\\
\midrule
Subsampled & 0.060 & 0.058 & 0.064 \\
$\ell_1$-CS & 1.160 & 0.630 & 1.471 \\
Standard 3D ALOHA & 0.989 & 0.936 & 1.062 \\
Proposed MC-Hankel & 1.160 & 1.083 & 1.251 \\
\bottomrule
\end{tabular}
\end{table}

\subsection{Discussion and insights}\label{s:Disc}
The results support the central hypothesis that joint low-rank structure and Maxwell-model consistency are complementary. Standard ALOHA exploits shared support among current source components, but its completed vectors can leave the physically admissible subspace. MC-Hankel explicitly removes that degree of freedom. The improvement is largest when the standard solution has a comparatively large model residual, as in J1 and noisy low-rate conditions, but it remains measurable even for J2, where standard ALOHA is already accurate.

The model residual is a useful mechanistic ablation, but it is not sufficient by itself: any hard projection can reduce physical inconsistency while increasing image error if the assumed model is wrong. In the present matched-model experiments, the projection simultaneously reduces model residual, Rel.\(L_2\), and the primary image metrics. This alignment supports the use of the constraint for these sources. The mixed Dice$_{55}$ behavior in noisy J1 also cautions against relying on one threshold or one visual rendering.

The proposed method occupies a different point in the design space from recent optimization and learning approaches. Structured gradient and Newton-like methods primarily accelerate or stabilize low-rank Hankel recovery \cite{CaiCaiYou2023,CaiHuangLuYou2025}; Hankel tensor completion extends the model to multiple measurements \cite{LiZhangWuCui2025}; and neural inverse source or operator methods learn a data-driven inverse map \cite{WiLeeOllerFazeli2025,ChenChangGuoWang2026,DongSuLiuChenChen2026}. MC-Hankel instead contributes a deterministic source-model projector that can, in principle, be combined with faster factor updates, tensor liftings, or learned priors.

Several limitations define the scope of the conclusions. First, the experiments are simulated and use only two source families. Second, the polarization is assumed known, and the same source decomposition is used for data generation and reconstruction. Third, the noisy study uses ideal circular complex AWGN and does not include sensor-position error, calibration drift, nonuniform frequency loss, or forward-model mismatch. Fourth, ranks are source-specific and fixed rather than selected automatically. Fifth, the full statistical analysis contains only five paired trials; the consistently positive bootstrap intervals and large effects are encouraging, but exact hypothesis tests remain underpowered. Finally, the reference is the fully sampled band-limited reconstruction, not the untruncated continuous source.

Future work should evaluate polarization perturbations, unmodeled source components, nonuniform spectral masks, experimental measurements, automatic rank selection, and larger Monte Carlo studies. The relaxation parameter in Eq.~\eqref{eq:relaxation} offers a direct mechanism for handling model mismatch: \(\gamma=0\) recovers standard ALOHA, while intermediate values trade strict consistency for robustness.

\section{Conclusion}\label{s:Conc}
A Maxwell-model-consistent joint 3D structured-Hankel method (MC-Hankel) has been developed for sparse multi-frequency electromagnetic inverse source reconstruction. By alternating low-rank completion with an exact source-subspace projection, conjugate symmetry, and noise-aware data consistency, the method preserves the shared-support advantage of multichannel ALOHA while eliminating nonphysical current coefficients. Across two source models, three sampling rates, clean and noisy conditions, and five paired trials, MC-Hankel improves every mean PSNR and SSIM result over standard ALOHA, reduces full-volume relative error, and drives the Maxwell-model residual to numerical precision with modest runtime overhead. The results establish a strong proof of concept; broader validation and a larger paired study are required before making definitive statistical claims.

\section*{Data and code availability}
Clean MATLAB project contains the J1 and J2 Fourier--Maxwell data, all reconstruction functions, validation scripts, statistical analysis, and figure/table generation code required to reproduce the reported experiments are available on Author's Github repository: \url{https://github.com/Shujaat123/MC-HANKEL}.

\section*{Competing interests}
The authors declared no potential conflicts of interest with respect to the research, authorship, and/or publication of this article.

\bibliographystyle{IEEEtran}

\begin{thebibliography}{10}
\providecommand{\url}[1]{#1}
\csname url@samestyle\endcsname
\providecommand{\newblock}{\relax}
\providecommand{\bibinfo}[2]{#2}
\providecommand{\BIBentrySTDinterwordspacing}{\spaceskip=0pt\relax}
\providecommand{\BIBentryALTinterwordstretchfactor}{4}
\providecommand{\BIBentryALTinterwordspacing}{\spaceskip=\fontdimen2\font plus
\BIBentryALTinterwordstretchfactor\fontdimen3\font minus
  \fontdimen4\font\relax}
\providecommand{\BIBforeignlanguage}[2]{{%
\expandafter\ifx\csname l@#1\endcsname\relax
\typeout{** WARNING: IEEEtran.bst: No hyphenation pattern has been}%
\typeout{** loaded for the language `#1'. Using the pattern for}%
\typeout{** the default language instead.}%
\else
\language=\csname l@#1\endcsname
\fi
#2}}
\providecommand{\BIBdecl}{\relax}
\BIBdecl

\bibitem{LeoneMaistoPierri2018}
G.~Leone, M.~A. Maisto, and R.~Pierri, ``Application of inverse source
  reconstruction to conformal antennas synthesis,'' \emph{IEEE Transactions on
  Antennas and Propagation}, vol.~66, no.~3, pp. 1436--1445, 2018.

\bibitem{Thio23}
B.~J. Thio, A.~S. Aberra, G.~E. Dessert, and W.~M. Grill, ``Ideal current
  dipoles are appropriate source representations for simulating neurons for
  intracranial recordings,'' \emph{Clinical Neurophysiology}, vol. 145, pp.
  26--35, 2023.

\bibitem{Beltrachini21}
L.~Beltrachini, N.~von Ellenrieder, R.~Eichardt, and J.~Haueisen, ``Optimal
  design of on-scalp electromagnetic sensor arrays for brain source
  localisation,'' \emph{Human Brain Mapping}, vol.~42, no.~15, pp. 4869--4879,
  2021.

\bibitem{Jerbi04}
K.~Jerbi, S.~Baillet, J.~Mosher, G.~Nolte, L.~Garnero, and R.~Leahy,
  ``Localization of realistic cortical activity in {MEG} using current
  multipoles,'' \emph{NeuroImage}, vol.~22, no.~2, pp. 779--793, 2004.

\bibitem{Takahashi22}
S.~Takahashi, K.~Suzuki, T.~Hanabusa, and S.~Kidera, ``Microwave subsurface
  imaging method by incorporating radar and tomographic approaches,''
  \emph{IEEE Transactions on Antennas and Propagation}, vol.~70, no.~11, pp.
  11\,009--11\,023, 2022.

\bibitem{AmmariN}
H.~Ammari, E.~Bretin, J.~Garnier, and A.~Wahab, ``Noise source localization in
  an attenuating medium,'' \emph{SIAM Journal on Applied Mathematics}, vol.~72,
  no.~1, pp. 317--336, 2012.

\bibitem{AmmariBaoFleming2002}
H.~Ammari, G.~Bao, and J.~L. Fleming, ``An inverse source problem for
  {M}axwell's equations in magnetoencephalography,'' \emph{SIAM Journal on
  Applied Mathematics}, vol.~62, no.~4, pp. 1369--1382, 2002.

\bibitem{Albanese06}
R.~Albanese and P.~B. Monk, ``The inverse source problem for {M}axwell's
  equations,'' \emph{Inverse Problems}, vol.~22, no.~3, p. 1023, 2006.

\bibitem{ISPs}
V.~Isakov, \emph{Inverse source problems}, ser. Mathematical surveys and
  monographs.\hskip 1em plus 0.5em minus 0.4em\relax Providence, R.I.: American
  Mathematical Society, 1990, vol.~34.

\bibitem{Valdivia12}
N.~P. Valdivia, ``Electromagnetic source identification using multiple
  frequency information,'' \emph{Inverse Problems}, vol.~28, no.~11, p. 115002,
  2012.

\bibitem{WAHAB}
A.~Wahab, A.~Rasheed, R.~Nawaz, and S.~Anjum, ``Localization of extended
  current source with finite frequencies,'' \emph{Comptes Rendus Mathematique},
  vol. 352, no.~11, pp. 917--921, 2014.

\bibitem{Bleistein}
N.~Bleistein and J.~K. Cohen, ``Nonuniqueness in the inverse source problem in
  acoustics and electromagnetics,'' \emph{Journal of Mathematical Physics},
  vol.~18, no.~2, pp. 194--201, 1977.

\bibitem{WangEM2019}
X.~Wang, M.~Song, Y.~Guo, H.~Li, and H.~Liu, ``Fourier method for identifying
  electromagnetic sources with multi-frequency far-field data,'' \emph{Journal
  of Computational and Applied Mathematics}, vol. 358, pp. 279--292, 2019.

\bibitem{vetterli2002sampling}
M.~Vetterli, P.~Marziliano, and T.~Blu, ``Sampling signals with finite rate of
  innovation,'' \emph{IEEE Transactions on Signal Processing}, vol.~50, no.~6,
  pp. 1417--1428, 2002.

\bibitem{ChenChi2014}
Y.~Chen and Y.~Chi, ``Robust spectral compressed sensing via structured matrix
  completion,'' \emph{IEEE Transactions on Information Theory}, vol.~60,
  no.~10, pp. 6576--6601, 2014.

\bibitem{JinYe2015}
K.~H. Jin and J.~C. Ye, ``Annihilating filter-based low-rank {H}ankel matrix
  approach for image inpainting,'' \emph{IEEE Transactions on Image
  Processing}, vol.~24, no.~11, pp. 3498--3511, 2015.

\bibitem{JinLeeYe2016}
K.~H. Jin, D.~Lee, and J.~C. Ye, ``A general gramework for compressed sensing
  and parallel {MRI} using annihilating filter based low-rank {H}ankel
  matrix,'' \emph{IEEE Transactions on Computational Imaging}, vol.~2, no.~4,
  pp. 480--495, 2016.

\bibitem{JacobManiYe2020}
M.~Jacob, M.~P. Mani, and J.~C. Ye, ``Structured low-rank algorithms: Theory,
  magnetic resonance applications, and links to machine learning,'' \emph{IEEE
  Signal Processing Magazine}, vol.~37, no.~1, pp. 54--68, 2020.

\bibitem{GuoWahabWang2023}
Y.~Guo, S.~Khan, A.~Wahab, and X.~Wang, ``Multipolar acoustic source
  reconstruction from sparse far-field data using {ALOHA},'' \emph{IEEE Signal
  Processing Letters}, vol.~30, pp. 1627--1631, 2023.

\bibitem{arxiv}
\BIBentryALTinterwordspacing
A.~K. Al-Shaqsi, H.~M. Al-Subhi, X.~Wang, S.~Khan, and A.~Wahab,
  ``Multi-frequency far-field data enrichment for electromagnetic source
  reconstruction,'' 2026. [Online]. Available:
  \url{https://arxiv.org/abs/2608.04829}
\BIBentrySTDinterwordspacing

\bibitem{CaiCaiYou2023}
H.~Cai, J.-F. Cai, and J.~You, ``Structured gradient descent for fast robust
  low-rank {H}ankel matrix completion,'' \emph{SIAM Journal on Scientific
  Computing}, vol.~45, no.~3, pp. A1172--A1198, 2023.

\bibitem{WuYangXu2024}
X.~Wu, Z.~Yang, and Z.~Xu, ``Multichannel frequency estimation with constant
  amplitude via convex structured low-rank approximation,'' \emph{SIAM Journal
  on Matrix Analysis and Applications}, vol.~45, no.~3, pp. 1643--1668, 2024.

\bibitem{CaiHuangLuYou2025}
H.~Cai, L.~Huang, X.~Lu, and J.~You, ``Accelerating ill-conditioned {H}ankel
  matrix recovery via structured {N}ewton-like descent,'' \emph{Inverse
  Problems}, vol.~41, no.~7, p. 075015, 2025.

\bibitem{LiZhangWuCui2025}
J.~Li, X.~Zhang, S.~Wu, and W.~Cui, ``Fast and provable {H}ankel tensor
  completion for multi-measurement spectral compressed sensing,'' \emph{IEEE
  Transactions on Signal Processing}, vol.~73, pp. 4060--4076, 2025.

\bibitem{WiLeeOllerFazeli2025}
Y.~Wi, J.~Lee, M.~Oller, and N.~Fazeli, ``Neural inverse source problem,'' in
  \emph{Proceedings of the 8th Conference on Robot Learning}, ser. Proceedings
  of Machine Learning Research, vol. 270.\hskip 1em plus 0.5em minus
  0.4em\relax PMLR, 2025, pp. 4371--4391.

\bibitem{ChenChangGuoWang2026}
H.~Chen, Y.~Chang, Y.~Guo, and Y.~Wang, ``A deep learning-enhanced {F}ourier
  method for the multi-frequency inverse source problem with sparse far-field
  data,'' \emph{arXiv preprint arXiv:2601.00427}, 2026.

\bibitem{DongSuLiuChenChen2026}
Q.~C. Dong, Z.-X. Su, Q.~H. Liu, W.~Chen, and Z.~Chen, ``Physics-informed
  neural operator for electromagnetic inverse scattering problems,''
  \emph{arXiv preprint arXiv:2603.25404}, 2026.

\bibitem{JiLiu2020}
X.~Ji and X.~Liu, ``Inverse electromagnetic source scattering problems with
  multifrequency sparse phased and phaseless far field data,'' \emph{SIAM
  Journal on Scientific Computing}, vol.~41, no.~6, pp. B1368--B1388, 2019.

\bibitem{GriesmaierSchmiedecke2017}
R.~Griesmaier and C.~Schmiedecke, ``A factorization method for multifrequency
  inverse source problems with sparse far field measurements,'' \emph{SIAM
  Journal on Imaging Sciences}, vol.~10, no.~4, pp. 2119--2139, 2017.

\bibitem{LiLiu2023}
J.~Li and X.~Liu, ``Reconstruction of multiscale electromagnetic sources from
  multifrequency electric far field patterns at sparse observation
  directions,'' \emph{Multiscale Modeling \& Simulation}, vol.~21, no.~2, pp.
  753--775, 2023.

\bibitem{Colton-Kress}
D.~Colton and R.~Kress, \emph{Inverse Acoustic and Electromagnetic Scattering
  Theory}, 4th~ed., ser. Applied Mathematical Sciences.\hskip 1em plus 0.5em
  minus 0.4em\relax Springer-Cham, 2019, vol.~93.

\bibitem{Lindell}
I.~Lindell, ``{TE}/{TM} decomposition of electromagnetic sources,'' \emph{IEEE
  Transactions on Antennas and Propagation}, vol.~36, no.~10, pp. 1382--1388,
  1988.

\bibitem{ye2016compressive}
J.~C. Ye, J.~M. Kim, K.~H. Jin, and K.~Lee, ``Compressive sampling using
  annihilating filter-based low-rank interpolation,'' \emph{IEEE Transactions
  on Information Theory}, vol.~63, no.~2, pp. 777--801, 2017.

\bibitem{FazelPongSunTseng2013}
M.~Fazel, T.~K. Pong, D.~Sun, and P.~Tseng, ``Hankel matrix rank minimization
  with applications to system identification and realization,'' \emph{SIAM
  Journal on Matrix Analysis and Applications}, vol.~34, no.~3, pp. 946--977,
  2013.

\end{thebibliography}

\end{document}